\documentclass[11pt,amsmath,amssymb]{article}
\usepackage{amsmath}

\usepackage[all]{xy}
\usepackage{amssymb}

\newcommand{\na}{\nabla}
\newcommand{\beq}{\begin{equation}}
\newcommand{\eeq}{\end{equation}}
\newcommand{\bit}{\begin{itemize}}
\newcommand{\eit}{\end{itemize}}
\newcommand{\ben}{\begin{enumerate}}
\newcommand{\een}{\end{enumerate}}
\newcommand{\la}{\langle}
\newcommand{\ra}{\rangle}

\newcommand{\fh}{{\mathfrak{H}}}

\newcommand{\bs}{\boldsymbol}

\newcommand{\mb}{\mbox}

\begin{document}
\begin{titlepage}

\begin{flushright}
\today
\end{flushright}

\vspace{1in}

\begin{center}

{\bf Modeling M-Theory Vacua via Gauged S-Duality}

\vspace{1in}

\normalsize

{Eiji Konishi\footnote{E-mail address: konishi.eiji@s04.mbox.media.kyoto-u.ac.jp}}

\normalsize
\vspace{.5in}

 {\em Faculty of Science, Kyoto University, Kyoto 606-8502, Japan}

\end{center}

\vspace{1in}

\baselineskip=24pt
\begin{abstract}We construct a model of M-theory vacua using gauged S-duality and the Chan-Paton symmetries by introducing an infinite number of open string charges. In the Bechi-Rouet-Stora-Tyutin formalism, the local description of the gauged S-duality on its moduli space of vacua is fully determined by one physical state condition on the vacua.
We introduce the string probe of the spatial degrees of freedom and define the increment of the cosmic time. The dimensionality of space-time and the gauge group of the low energy effective theory originate in the symmetries (with or without their breakdown) in our model. This modeling leads to the derived category formulation of the quantum mechanical world including gravity and to the concept of a non-linear potential of gauged and affinized S-duality which specifies the morphism structure of this derived category.
\end{abstract}

\vspace{.7in}
 
\end{titlepage}
\section{{Introduction}}

A self-consistent unifying quantum theory of the fundamental forces of Nature including gravity has been sought by elementary particle physicists for many decades. String theory is currently thought to be a promising candidate for such a theory. The first string theory revolution occurred in 1984 following M. B. Green and J. H. Schwarz's celebrated discoveries.\cite{GS1,GS2,GSW} Since then it has been formulated to be the present five traditional forms and, sharing its position with loop quantum gravity\cite{loop1,loop2,loop3,loop4,loop5}, is regarded as a candidate for the quantum theory of gravity.
Since the second revolution of string theory occurred around 1995, we have seen that the five traditional ten-dimensional string theories (of type I, type IIA, type IIB, heterotic $E_8\times E_8$ and heterotic $SO(32)$) can be unified in an eleven-dimensional {\it{M-theory}} which appears as the strong coupling limit of type IIA string theory with its Kaluza-Klein modes of Dirichlet particles (D-particles) and each string theory describes a different aspect of the same theory.\cite{WP1,WP2,WP3} In this sense, throughout this paper, we use the term {\it{M-theory vacua}} to mean the vacua underlying not only M-theory but also the five string theories. The low energy effective theory of the M-theory is eleven-dimensional ${\cal{N}}=1$ supergravity. Due to the eleven-dimensional ${\cal{N}}=1$ supersymmetry algebra, it is recognized that the fundamental and dynamical ingredients of M-theory are the M2-branes and M5-branes admitted as the Bogomol'nyi-Prasad-Sommerfield (BPS) solutions, the D-particles and Kaluza-Klein monopoles, that is, D6-branes. Recent research has revealed that an infinite-dimensional representation of the Nambu bracket of the gauge symmetries of the Lie-3 algebra of the field theory on the world volume of the infinitely multiplied M2-branes is a field theory on the world volume of an M5-brane, so it is expected that M2-branes are more fundamental ingredients than M5-branes.\cite{BL1,BL2,BL3} The fundamental strings (F-strings) and D2-branes in type IIA string theory are identified with M2-branes that are, respectively, wrapped and unwrapped around the eleventh-dimension. After the discovery of M-theory, various novel non-perturbative (out of perturbative schemes) formulations of it, such as M(atrix)-theories, were proposed in addition to the string field theories that had been studied for a decade before the second revolution.\cite{HIKKO,Witten,BFSS,IKKT,Taylor}

In M-theory, the most important unsolved issue is to deduce the {\it{true}} vacuum. In the perturbative formulation of string theory, the potential is flat for infinitely many {\it{false}} vacua. Usually, in Kaluza-Klein reduction, one considers four-dimensional compactification with an internal Calabi-Yau 3-fold which has the $SU(3)$ holonomy group and retains ${\cal{N}}=1$ supersymmetry on the non-compactified four dimensions. A decade ago, string theorists introduced the flux compactifications in addition to these standard Calabi-Yau compactifications and stabilized the moduli appearing in the theory by minimizing the potential of the internal fluxes\cite{GKP} and a huge number of realistic de-Sitter vacua have been constructed in the string landscape.\cite{KKLT,frev1,frev2,frev3} The motivation of this paper is to propose a novel non-perturbative formulation of M-theory vacua, which guides us to reduce the number of vacua and address the non-perturbative properties of the reduced vacua.

We begin by explaining the theory of gauged strong-weak coupling duality (gauged S-duality) expected to produce such a formulation of M-theory vacua. The links between the five traditional string theories and M-theory are the string dualities\cite{T1,T2,r1,r2,r3}, which are classified into two kinds. The first kind is the S-duality which relates the strong and weak coupling
 phases of the same theory or of two different theories. Type IIB string theory is an example of the former case;
on the other hand, a familiar example of the latter case is
the heterotic string theory with the $SO(32)$ gauge group which is S-dual to type I string theory with same gauge group in $D=10$. The second kind is the target space duality, T-duality, which is examined perturbatively. A simple example is a bosonic closed string theory whose one spatial coordinate is compactified 
on a circle with radius $R$. The 
perturbative spectrum of this theory matches with the one whose 
corresponding spatial coordinate is compactified on a circle with 
radius $1/R$. This is a consequence of the modular invariance of the partition function under the exchange of the temporal and the string-coordinate directions on F-string world sheets (i.e., the exchange of winding and unwinding strings around the circle). The T-duality translates type IIA and type IIB theories into each other in this way and shifts the dimensions of D-branes by plus and minus one. Using combinations of S-duality and T-duality with compactifications, there is a duality web between all of the five traditional ten-dimensional string theories and the eleven-dimensional M-theory.

The notion of gauged S-duality, which was recently introduced by the author\cite{K}, has a representation on M-theory vacua parameterized by the coupling constant $g_s$. Illustratively, in type IIB supergravity, usually, S-duality symmetry is considered as a non-linear ${SL}(2,{\boldsymbol{R}})_S$ {{global}} symmetry on the Poincar\'e upper half-plane $\fh$ of the coupling constant.\cite{r1} In contrast, we take the gauge transformation on each vacuum on $\fh$ to be independent of the others and consider the linear ${SL}(2,{\boldsymbol{R}})_S$ local symmetry for the axion-dilaton moduli matrix ${\mathcal{M}}$ generated by the infinitesimal gauge transformations\begin{equation}{{\mathcal{M}}}=\left(\begin{array}{cc}\chi^2e^\Phi+e^{-\Phi}&\chi e^\Phi\\\chi e^\Phi&e^\Phi\end{array}\right)\;,\ \ \delta {{\mathcal{M}}}= \epsilon({\Sigma}{{\mathcal{M}}}+{{\mathcal{M}}}{\Sigma}^{T})\;,\label{eq:tra}\end{equation} where the matrices ${\Sigma}$ are a basis of ${sl}(2,{\boldsymbol{R}})_S$ and $\epsilon$ depends on the coupling constant $g_s$.
In string theory, F-strings and D-branes are the physical states associated with the gauged S-duality. As already mentioned, in type IIA/M-theory, D-particles are the Kaluza-Klein particles of the eleventh dimension whose radius is reciprocal to the coupling constant.\cite{WP1,WP2,WP3,BFSS} Their cousins in type IIB string theory, the D-strings and fundamental open strings, form a doublet of the S-duality symmetry via a coupling to the moduli matrix.

An introductory study of gauged S-duality in type IIA/M theory was given by the author' previous paper.\cite{K}
In the present investigation of gauged S-duality, by utilizing the S-duality doublet of axion and dilaton and that of F- and D-strings in type IIB string theory, we formulate type IIB aspect of M-theory vacua under the moduli of string dualities.
As will be explained shortly, we introduce the affinization of S-duality (i.e., to incorporate the world sheet degrees of freedom (d.o.f.) of a perturbative string theory into an affine Lie algebra based on $sl(2,{\boldsymbol{R}})_S$ algebra) in addition to the gauging of S-duality.
Our modeling of M-theory vacua by affinized and gauged S-duality contains T-duality in the weak string coupling region as the invariance of the vacua under the modular transformations of the modulus parameter (see Eq.(\ref{eq:modular})). Hereafter we refer to affinized and gauged S-duality simply as {\it{gauged S-duality}} and consider type IIB string theory.

Here, we make three remarks. First, the process of gauging S- and T- dualities is applicable only to the moduli space of vacua (to be defined later) and not to the field theory of each individual vacuum. Obviously, the S- and T-duality gauge equivalent class of each vacuum on the moduli space corresponds to a solution of the equations of motion of the S-duality invariant action of type IIB string theory under the moduli of S- and T-dualities. Hence, the special vacuum configurations derived from the S-duality invariant action are the local descriptions of the gauged S-duality on its moduli space. In this sense, the field theory of gauged S-duality on the moduli space of vacua, treated in Section 2, can be seen as a generalization of type IIB string theory.
Second, gauging S-duality removes the artificial distinction between the excitations of F- and D-strings. The field theory of gauged S-duality has the enlarged Hilbert space of the third quantized D-brane fields instead of the one of the second quantized string fields.
Third, the physical substance of gauging process is to regard various S-duality gauge bosons as the excitation modes of F-strings by the Chan-Paton modes on D-strings as seen in Yoneya's D-brane field theory, which is a {\it{theory of anything}}.\cite{Y1,Y2} 
The task for us is to extract the {\it{theory of everything}} from {\it{theories of anything}}.

We state the reason why we gauge S-duality and relate it to the perturbative string symmetries. 

{{First, we explain why we gauge S-duality according to Yoneya's paper.\cite{Yoneya0} To do this, we consider an analogy between the duality in string field theories and the Coleman-Mandelstam duality (CM duality) in two-dimensional space-time.\cite{CM1,CM2} CM duality relates between the two dimensional fermionic system of the massive Thirring model and the solitonic solutions (kinks) of the bosonic sine-Gordon model. Here, we consider the second quantized theory of the sine-Gordon model. In string theory, D-branes are non-trivial classical kink solutions of the supergravity approximation of the closed string field theory. In the analogy between string theory and CM duality, the closed string field theory corresponds to the sine-Gordon field.  CM duality explains the duality between open string field theory, that is, as a massive Thirring model and closed string field theory, that is, as a sine-Gordon model. When we second-quantize the D-brane system, it is recognized to be a second-quantized open string field theory. This open-closed string duality explains the duality between closed string field theory and the open string field degrees of freedom in D-brane field theory. Thus, the dualities between these three string field theories are explained. We shall identify the physical state of the gauge theory of S-duality with the vacuum of type IIB string theory. Accordingly, when we consider the second quantized M-theory, we are inevitably led to the gauging of S-duality in order to exclude the artificial distinctions between perturbative and non-perturbative excitations of strings in their field theories. 

Now, we explain why we affinize the gauged S-duality.
In string theory, besides the S-duality symmetry there are Chan-Paton gauge symmetries, whose generators are as many as the number of degrees of freedom. Following the formulation of the type IIB matrix model\cite{IKKT}, to cover all of the degrees of freedom, we combine them in a manner to be explained later (we call this process {\it{affinization}}).}}

The purpose of this paper is to model type IIB string theory including its non-perturbative dynamics. The following model has two distinct structures. (As seen in the following, they are not achieved till we consider the gauge theory of duality symmetries.) First structure is the gauge field theory on the moduli space of vacua detailed in Section 2.
Each vacuum is specified by the Kugo-Ojima physical state condition\cite{KO1,KO2} and is the stable field configuration when we regard the Bechi, Rouet, Stora and Tyutin (BRST) charge\cite{BRST1,BRST2,BRST3,BRST4} as a differential. As will be explained, this gauge field theory is just an infinitesimal local description of the moduli space of vacua in type IIB string theory. Non-perturbative effects are not yet described and non-perturbative field configurations are fixed and not dynamical. So, the contents of this stage of the modeling are not so different from those of a perturbative string theory. The non-perturbative description or dynamics of type IIB string theory, i.e., transition between the stable configurations, is achieved by introducing another non-linear potential, as the second structure of the model, which can describe the moduli space of vacua globally. Then, we can describe non-perturbative effects, such as an infinite many body effect and the dynamics of D-branes. The way to introduce the second gauge potential is based on the derived category structure of the state spaces generated by a fixed vacuum. This derived category structure bases on Eq.(\ref{eq:time1}) and results from the perturbative string symmetry, T-duality. Due to these structures, the theory studied in this paper is, on the whole, equivalent to the standard non-perturbative formulation of type IIB string theory\cite{IKKT} including the issue of the unitarity.
{{Here, we note that due to the non-compactness of $SL(2,{\bs{R}})$, it has no finite dimensional unitary representation. Thus, throughout this paper we consider its infinite dimensional unitary representation.}}

On the basis of the idea of gauging S-duality, we construct a model of type IIB string theory vacua with the conserved charges of its full symmetries and its time development deformation variables (we call them {\it{time variables}}). 

Our model is based on the Neveu, Schwarz and Ramond (NSR) model of type IIB string theory that contains the massless and bosonic excitations of the axion $\hat{\chi}$, dilaton $\hat{\Phi}$, graviton $\hat{g}_{MN}$, 2-form Neveu-Schwarz-Neveu-Schwarz (NS-NS) and Ramond-Ramond (R-R) potentials ($\hat{B}^{(i)}_{MN}$ for $i=1,2$, respectively) and the R-R 4-form potential with its self-dual field strength. A hat indicates that a field is ten-dimensional. The effective action in the Einstein frame is\cite{hull,jhs}
 \begin{eqnarray}
S=\frac{1}{2\kappa^2}\int d^{10}x\sqrt{-\hat{g}}
\biggl[\hat{R}_{\hat{g}}+\frac{1}{4}{\rm{Tr}}(\partial_M\hat{{\cal{M}}}\partial^M\hat{{\cal{M}}}^{-1})-\frac{1}{12}{\hat{{\boldsymbol{H}}}}_{MNP}^T\hat{{\cal{M}}}{\hat{{\boldsymbol{H}}}}^{MNP}\biggr]\;, \label{eq:NSR}
\end{eqnarray}
where the vector of $H$-fields is
\begin{equation}{{\hat{{\boldsymbol{H}}}}}_{MNP}=\left(\begin{array}{cc}\hat{H}^{(1)}\\ \hat{H}^{(2)}\end{array}\right)_{MNP}\;,
\end{equation}
and $\hat{H}^{(i)}=d\hat{B}^{(i)}$.
We exclude the R-R 4-form potential from consideration.
The action in Eq.(\ref{eq:NSR}) is manifestly invariant under the S-duality transformations
\begin{eqnarray}\hat{{\cal{M}}}\to\Lambda\hat{{\cal{M}}}\Lambda^T\;,\ \ {\hat{{\boldsymbol{H}}}}\to (\Lambda^T)^{-1}{\hat{{\boldsymbol{H}}}}\;,\ \ 
\hat{g}_{MN}\to \hat{g}_{MN}\;,\label{eq:Sgroup}
\end{eqnarray}where $\Lambda\in SL(2,{\boldsymbol{R}})_S$.

In the following, we gauge and quantize the S-duality group of Eq.(\ref{eq:Sgroup}).

We regard the pair of the axion and dilaton and that of F- and D-strings as the gauge bosons of gauged S-duality. Since the axion and dilaton parameterize the coset $SL(2,{\bs{R}})/SO(2)\simeq{\fh}$, the corresponding gauge potentials, as the connections on the fiber bundle, satisfy\begin{equation}a_n\in d{\fh}\;,\end{equation}
for the tangent space of the Poincar\'e upper half plane $d{\fh}$ and index $n$ of the base space coordinates $s_n$. 
However, we assume that the gauge group is originally $SL(2,{\bs{R}})_S$ and the Lie algebraic constraints on the field operators (see Eq.(\ref{eq:gauge})) reduce it to the coset $SL(2,{\bs{R}})_S/SO(2)$. Then, the number of the generators of the gauge symmetry is still three. We denote the generators of $SL(2,{\bs{R}})_S$ by $Q^i$ for $i=0,1,2$. We define the two-dimensional representations\footnote{In the following, we refer to the representations of generators merely as the {\it{generators}}.} of the gauge symmetry, $\Sigma^i$ for $i=0,1,2$, such that the field variables $\psi_r$ satisfy
\begin{equation}[\psi_r,Q^i]=(\Sigma^i)_{rs}\psi_s\;,\ \ r,s=1,2\;.\label{eq:trans1}\end{equation}
 The infinite dimensional unitary representations of the gauge transformations corresponding to Eq.(\ref{eq:trans1})
\begin{equation}{\cal{U}}(\lambda(s))=\exp\Biggl(i\sum_{i=0}^{2} \Sigma^i\lambda^i(s)\Biggr)\;,\end{equation} act on the field variables $\psi$ and the gauge potential $a_n$ as\footnote{The partial differentials $\partial/\partial s_n$ will be written as $\partial_n$.}\begin{equation}\psi\to {\cal{U}}(\lambda(s))\psi\;,\ \ a_n\to {\cal{U}}(\lambda(s)) a_n{\cal{U}}(\lambda(s))^{-1}-\frac{1}{ig}{\cal{U}}(\lambda(s))\partial_n {\cal{U}}(\lambda(s))^{-1}\;,\label{eq:trans2}\end{equation} where $g$ is the coupling constant.
In type IIB string theory, the S-duality gauge group of Eq.(\ref{eq:Sgroup}) is quantized from $SL(2,{\bs{R}})_S$ to $SL(2,{\bs{Z}})_S$ by imposing Dirac's charge quantization condition on the charges $\Sigma^i\lambda^i$ for $i=0,1,2$. In this paper our scheme for this quantization process is, first, to build the theory for the continuous family $\lambda(s)$ (in the BRST transformation this is the ghost field); second, we restrict the coordinates $s_n$ to be discrete, so that the transformation operators ${\cal{U}}(\lambda(s))$ belong to the representation of $SL(2,{\bs{Z}})_S$. However, this restriction of the coordinates $s_n$ does not hinder the mathematical structure of our theory before the restriction and it will be sufficient only to refer to the necessity of the restriction here. When we refer to the coordinates $s_n$ on the base space in type IIB string theory (not in type IIB supergravity), we regard them as discrete variables.

In our modeling we, first, introduce the variable $t_0$ and its differential $q_0={\partial}/{\partial t_0}$ to represent the infinitesimal S-duality transformations in $sl(2,{\boldsymbol{R}})_{{{{S}}}}$ with the infinitesimal generators
\begin{equation}
sl(2,{\bs{R}})_{{{{S}}}}=\la\Sigma^i(0);\ i=0,1,2\ra_{{\bs{R}}}\;.\label{eq:BPS}
\end{equation} In addition, we introduce canonical time variables ${t}_n$, where $n\in{\boldsymbol{Z}}\backslash\{0\}$, for the open string charges ${{q}}_n$ associated with independent gauge symmetries of the Chan-Paton factors (the d.o.f. of the coupling edges of open strings to a 1-form on D-strings, for $n>0$, and anti-D-strings, for $n<0$). Thus the gauge bosons, which couple to the string states with the Chan-Paton charges, correspond to the Chan-Paton gauge bosons and their anti-bosons with the gauge group $\bigoplus_{N\ge1}U(N)$.
{{Here, we note that the unitary group $U(N)$ has $N^2$ generators. For our purpose we focus on the multiplicities of D-branes in the vacua, so we simplify the situation by ignoring the internal degrees of freedom of D-branes, which are those of the bound states of D-branes and open strings, and represent the $N^2$ generators of $U(N)$ by one generator $q_n$, satisfying an affine Lie algebra.}}
By introducing this infinite number of open string charges, via Noether's theorem, we affinize the $sl(2,{\boldsymbol{R}})_{{{{S}}}}$ generators of the gauge transformation in order to create and annihilate F-strings, D-strings and anti-D-strings.\cite{Y1,Y2}
The affinization to the $\hat{\mathfrak{g}}=\widehat{sl(2,{\boldsymbol{R}})_{{{{S}}}}}$ algebra is given by the affine algebra of the generators\begin{subequations}
\begin{align}
&[\Sigma^1(l_1),\Sigma^2(l_2)]_-=[\Sigma^1,\Sigma^2]_-(l_1+l_2)+l_1\delta_{l_1,-l_2}(\Sigma^1,\Sigma^2)z\;,\label{eq:affine}\\
&[z,\hat{{\mathfrak{g}}}]_-=0\;,\label{eq:affine2}\\
&[\partial,\Sigma(l)]_-=l\Sigma(l)\;,\label{eq:affine3}
\end{align}
\end{subequations}for $l\in{\boldsymbol{Z}}$. In the present model, to also incorporate the BPS supersymmetry\cite{WP2} into the $\hat{{\mathfrak{g}}}$ gauge symmetries as a simplified correspondence between states by bosonization, we extend the moduli space of S-duality multiplets from the cosets $SL(2,{\bs{R}})/SO(2)$ of the axion and dilaton to $SL(2,{\bs{R}})$ by introducing the dilatino included in the NSR model of type IIB string theory. We denote the generator belonging to the Cartan subalgebra of Eq.(\ref{eq:BPS}), which survives in the S-duality group $SL(2,{\bs{R}})_S$ under the modulo of its maximal compact subgroup $SO(2)$, between two Cartan subalgebras of Eq.(\ref{eq:BPS}) by $\Sigma^0(0)$. Then, we interpret the generators in $\hat{{\mathfrak{g}}}$ for the $i=0,1,2$ parts to be the creation and annihilation operators of R-R D-strings, NS-NS F-strings and bosonized NS-R F-strings respectively. The $sl(2,{\boldsymbol{R}})_{{{{S}}}}$ symmetries on the coupling constant and the antisymmetric tensor parts of NS-NS and R-R states are realized by an $SO(2,1)_{{{{S}}}}$ rotation about the third axis (i.e., NS-R states), and the BPS supersymmetry between NS-NS and bosonized NS-R states is realized by the affine Weyl symmetry. Since the R-R D-strings are the BPS saturated states,\cite{WP2} we associate them with the axis fixed under the affine Weyl group action. Here, we treat the bosonization of the NS-R states in the Hilbert space of two-dimensional conformal field theory by truncating their fermionic Klein factor to adjust them to the bosonic Heisenberg generators in the affine Lie algebra. Thus, the BPS generators that we consider also are bosonic. In the following, we refer to these two processes merely as {\it{bosonization}}.

The rest of this paper is organized as follows. 

In the next section, we consider the infinitesimal local description of the moduli space of vacua under the Kugo-Ojima physical state condition when we regard the BRST charge as a differential. As our scheme, we first derive the Kugo-Ojima physical state condition of the gauged S-duality and introduce its solution in the Kac-Peterson form of the theta function; second, we introduce the string-probe of the spatial parts of the Utiyama gravitational gauge field\cite{U} and defining the increment of the cosmic time using the Casimir operator\cite{Casimir} of the representation of the affine Lie algebra in the field operators; third, we conjecture the equivalence between our wave function and the wave function of the Universe and describe the cosmic time developments of the wave functions of systems of microscopic or macroscopically coherent quantum fields.
We stress that in Section 2 we have not described the non-perturbative effects yet and non-perturbative field configurations are fixed and not dynamical. So, the contents of Section 2 are not so different from the modeling of a perturbative string theory. In Section 5, we generalize the results in Section 2 by introducing the non-linear potential which can describe the moduli space of vacua globally. Then, we can describe the non-perturbative effects, i.e., the dynamics of D-branes etc. This section is the edge of the logical consequences and also serves for the section of conclusion. In Section 6, we discuss the nature of the cosmic time in our model.
 In the Appendix, we formulate the renormalization of vacua by Whitham's multi-phase average method.\cite{Wh1,Wh2}

Throughout this paper, we denote the Lie bracket and the (anti-)commutator by $[,]$ and $[,]_-$ ($\{,\}$), respectively.
\section{Infinitesimal Description Using the BRST Field Theory}
\subsection{Preliminaries}
Our gauge theory is the $\hat{{\mathfrak{g}}}$ generalized Yang-Mills theory on an infinite-dimensional base space with coordinates $(s_n)_n$, where $n\in{\boldsymbol{Z}}$, and a Euclidean background metric. The coupling constant $g_s$ is introduced by a variable whose symmetry transformation property is covered by the variables of $H$-fields. Thus the dependences on the coupling constant $g_s$ will be suppressed in equations. We call the underlying principle bundle the {\it{moduli space of vacua}}. \footnote{
{{Here, the fiber space of the principle bundle is the corresponding infinite dimensional Lie group of $\widehat{sl(2,{\bs{R}})}_{{{{S}}}}$, that is, the Kac-Moody group.}}} For the $\hat{{\mathfrak{g}}}$-valued classical fields, the gauge potential $a_n$, the Faddeev-Popov ghost field $b$ and the anti-ghost field $\bar{b}$ are\begin{equation}a_n=\sum_{i=0}^2\sum_{l\in{\boldsymbol{Z}}}\Sigma^i(l)a_n^{i,l}\;,\ b=\sum_{i=0}^2\sum_{l\in{\boldsymbol{Z}}}\Sigma^i(l)b^{i,l}\;,\ 
\bar{b}=\sum_{i=0}^2\sum_{l\in{\boldsymbol{Z}}}\Sigma^i(l)\bar{b}^{i,l}\;.\end{equation} 

The Lagrangian is given by
\begin{eqnarray}
{\cal{L}}= -\frac{1}{4}F_{mn}^{i,l} F^{mn\ i,l}+\partial^n\bar{{b}}^{i,l}D_n{b}^{i,l}+\alpha\frac{\bigl(\gamma^{i,l}\bigr)^2}{2}
-i\partial^n\gamma^{i,l}{{a}}_n^{i,l}\;,\label{eq:YM}
\end{eqnarray}
where the field strength of the gauge potential is \begin{equation}-igF_{mn}=[D_m,D_n]_-\;,\end{equation} with covariant derivative \begin{equation}D_n\phi=\partial_n\phi+i{{g}}[a_n,\phi]\;,\end{equation} and its gauge is $\alpha$. 
We have introduced\begin{equation}\gamma=\partial_na_n\;,\ \ \bar{\gamma}=-\gamma+i{{g}}[b,\bar{b}]\;,\end{equation}
where the indices $m$ and $n$ are contracted and the $\hat{{\mathfrak{g}}}$ indices $i$ and $l$, which label the field component of the generator $\Sigma^i(l)$, are contracted until Eq.(\ref{eq:current}).

The classical constraints on the fields are
\begin{subequations}
\begin{align}
&D^m F_{mn}+\partial_n \gamma-i{{g}}[\partial_n \bar{b},{b}]=0\;,\label{eq:cgauge}\\
&\partial^nD_n{b}=0\;,\label{eq:cghost} \\
&D^n\partial_n\bar{b}=0\;,\label{eq:antighost}
\end{align}
\end{subequations}
for the gauge potential $a_n$, the ghost field $b$ and the anti-ghost field $\bar{b}$.

For any of the classical fields, which we generically label $\phi$, we define each field component $\phi^{i,l}$ so that the equations of motion give $\phi(s)$ from the field at $s=0$ through multiplication by an overall factor $U(s)$ (the coordinate conditions on the classical fields):
\begin{equation}
\phi(s)=U(s)\phi(0)\;,\ \ U(s)=\exp\Biggl(\sum_{n\in{\boldsymbol{Z}}}s_n{{ad}}(J_n)\Biggr)\;,\label{eq:s}
\end{equation}
where the physical meanings of the coordinates $s_n$ can be recognized as the time variables for the Virasoro-like operators $J_n$ for $n\in{\boldsymbol{Z}}$:
\begin{subequations}
\begin{align}
&[J_n,\Sigma^i(l)]_-=\Sigma^i(l+n)\;,\\
&[J_n,z]_-=0\;,\\
&[J_n,\partial]_-=-nJ_n\;,\\
&[J_m,J_n]_-=0\;.\label{eq:J}
\end{align}
\end{subequations}
In the NSR formalism the Virasoro operators are the coefficients of the mode expansion of the energy-momentum tensor on the F-string world sheet.
In Eq.(\ref{eq:s}) the $\phi^{i,l}(0)$ are just numbers.

 From the algebra in Eq.(\ref{eq:J}), it follows that 
\begin{equation}
\partial_n \phi^{i,l}(s)=\phi^{i,l-n}(s)\;.\label{eq:diff}
\end{equation}

Conversely, the coordinate dependence of the classical field is determined by Eq.(\ref{eq:diff}).
Thus it is easy to check the identities
 \begin{subequations}
 \begin{align}
\partial_n\phi(s)&=U(s)\partial_n \phi(0)\;,\label{eq:s2}\\
 [\phi_1,\phi_2](s)&=U(s)[\phi_1,\phi_2](0)\;.\label{eq:s3}
\end{align}
 \end{subequations}
 Thus the desired conditions on $\phi(s)$ are satisfied.

In our model we assume an infinite number of relations between the gauge potentials $a_n$:
\begin{equation}
\partial_na_n =\kappa_n a_0\;,\ \ \kappa_n=\kappa_\pm^{|n|}\ \ (\pm n\ge0)\;.\label{eq:Her}
\end{equation}
where the coordinate index $n\in{\boldsymbol{Z}}$ is not contracted, and the factors $\kappa_n$ are constant.

As discussed later, we also assume the relation \begin{equation}b=-\bar{b}\;.\label{eq:super2}\end{equation} Using Eqs.(\ref{eq:cghost}), (\ref{eq:antighost}), (\ref{eq:Her}) and (\ref{eq:super2}), it immediately follows that
\begin{equation}
[a_0,b]=0\;.\label{eq:relation}
\end{equation}

By fixing the gauge, we turn these fields (i.e., the gauge potential $a_n$, the ghost field $b$ and the anti-ghost field $\bar{b}$) into Hermitian operators\footnote{We will suppress the hat indicating the field operators in the equations without further notice.}
\begin{equation}\hat{\phi}(s)=\sum_{i=0}^2\sum_{{{l}}\in {\boldsymbol{Z}}}\Sigma^i({{l}})\int d{{{p}}}\,c_{\phi}^{i,l}(s,{{{p}}}){{{\Pi}}}_\phi^{i,l}({{p}})+({\mbox{Other\ components}})\;,\label{eq:gauge}\end{equation} where we denote the canonically conjugate variables (`momenta') of the variables of the following introduced representation space by $({{{p_n}}})_n$. (Eq.(\ref{eq:gauge}) is not a mode expansion, since Eq.(\ref{eq:YM}) contains nonlinear interactions.)
The coefficients $c_\phi$ are elements of the Fourier dual space of the representation space.
We denote the coefficients $c_{a_n}$, $c_b$ and $c_{\bar{b}}$ of Eq.(\ref{eq:gauge}) by $c_n$, $c$ and $\bar{c}$ respectively.
In Eq.(\ref{eq:gauge}), we introduced the Frenkel-Kac highest weight representations of the superalgebra\cite{VGKAC} of $\hat{{\mathfrak{g}}}$, which is constructed by adding the fermionic generators corresponding to the creation and annihilation operators of the ghost field and the anti-ghost field to Eq.(\ref{eq:affine}), on their common representation space ${{{\Pi}}}_\phi^{i,l}({p})$, where $i=0,1,2$ and $l\in{\boldsymbol{Z}}$.\cite{dual} We set\begin{equation}\Pi_{a_n}=\Pi_{a_0}\;,\ \ n\in{\boldsymbol{Z}}\;.\label{eq:gaugesub}\end{equation}
The two operators $\Pi_\phi(p)$ and $\Pi_\phi(p^\prime)$ satisfy the superalgebra of $\hat{{\mathfrak{g}}}$ apart from the overall factor $\delta(p-p^\prime)$ and are realized by applying the harmonic oscillator representations to both of the Heisenberg part and the S-triple part of $\hat{{\mathfrak{g}}}$ due to the fact that $sl(2,{\bs{R}})$ has the infinite dimensional representations written by creation and annihilation operators.
Here, the field components of $\phi$ need to satisfy the canonical commutation and anti-commutation relations of Eq.(\ref{eq:YM}) by a time variable as operators. This is the momentum condition on the coefficients $c_\phi$. The coordinate condition is still given by Eq.(\ref{eq:s}) on the field operators $\hat{\phi}$. Hereafter we refer to the superalgebra of $\hat{{\mathfrak{g}}}$ simply as {\it{$\hat{{\mathfrak{g}}}$ algebra}}.

The reason why the operator basis in Eq.(\ref{eq:gauge}) is described by a representation $\Pi_\phi$ and satisfies Eq.(\ref{eq:gaugesub}) is as follows.
Since, as explained in the Introduction, the field operators need to be under the Lie algebraic constraints, namely the operator basis ${\hat{O}}_\phi^{\Sigma}$ of the field operators $\hat{\phi}^\Sigma$ are tangent vectors on the section of the moduli space of vacua at the point in the fiber space specified by the field coefficients, they are written as linear combinations of the representations $\tilde{\Pi}_\phi^\Sigma$. The representations $\tilde{\Pi}_\phi^\Sigma$ are also the symmetric actions on the representation space. Thus the rings $R_\phi^\Sigma$ of the operator basis of the field operators ${\hat{O}}_\phi^\Sigma$ for the generators $\Sigma$ satisfy the filtration conditions for all pairs of generators. The unique gauge fixing conditions on the rings ${{R}}_\phi^{\Sigma}$ for the generators $\Sigma$, which are consistent with the filtration conditions, are the representations ${\Pi}_\phi^\Sigma$.

\subsection{BRST Field Theory}
\subsubsection{BRST and NO Symmetries}
On the basis of the preliminaries so far, we consider the two gauge symmetries of Eq.(\ref{eq:YM}) after the gauge fixing. The BRST and Nakanishi and Ojima (NO) symmetries are defined by the following infinitesimal gauge transformations with the gauge functions $b$ (BRST) and $\bar{b}$ (NO).\cite{BRST1,BRST2,BRST3,BRST4}
\begin{equation}
\begin{cases}
\delta^{(1)} a_n&=\epsilon D_n b\\ \delta^{(1)} b&=-\epsilon\frac{1}{2}g[b,b]\\ \delta^{(1)}\bar{b}&=i\epsilon\gamma\\ \delta^{(1)}\gamma&=0\;,\end{cases}\ \ \begin{cases}\delta^{(2)} a_n&=\epsilon D_n\bar{b}\\ \delta^{(2)}\bar{b}&=-\epsilon\frac{1}{2}{{g}}[\bar{b},\bar{b}]\\ \delta^{(2)}{b}&=i\epsilon\bar{\gamma}\\ \delta^{(2)}\bar{\gamma}&=0\;,\label{eq:BRST}\end{cases}\end{equation}
where the parameter $\epsilon$ is an anti-commuting $c$-number.
The transformations $\delta^{(1)}$ and $\delta^{(2)}$ are nilpotent.

As is well known, these transformations for the ghost and anti-ghost fields have a clear meaning in the geometry of the principle bundle.\cite{YM,ghost} Once the gauge d.o.f. is fixed, the gauge potential $a_n$ and the ghost field $b_{\alpha}$, for $\alpha=(i,l)$ with $i=0,1,2$ and $l\in{\boldsymbol{Z}}$, can be regarded as contravariant components of the Ehresmann vertical connection $\nabla_v$ on the section of the principle bundle:
\begin{equation}
\nabla_v =a_nds^n +b_{\alpha} dy^{\alpha}\;,
\end{equation}
where $y^{\alpha}$ are the coordinates of the internal fiber space. For later discussion, we put $dy^l=\sum_i dy^{i,l}$. Since any system of coordinates could be used without affecting the definitions of field operators, we distinguish between the index of these coordinates and the index of the $\hat{{\mathfrak{g}}}$ generators (i.e., the cotangent space index).\cite{ghost}
The BRST transformation is constructed from the Maurer-Cartan equations for the curvatures of $ig\nabla_v$\cite{ghost} \begin{equation}R_{\alpha\beta}(b,s,y)=R_{\alpha n}(a,b,s,y)=0\;,\end{equation}
 as
\begin{equation}\delta^{(1)}=d y^{\alpha} \partial_{\alpha}\;,\ \ (\delta^{(1)})^2=0\;,\end{equation}where we put $\partial_{\alpha}=\partial/\partial y^{\alpha}$ and two $dy^{\alpha}$ anti-commute.

Then
 \begin{equation}R_{\alpha\beta}(-b,\tilde{s},\tilde{y})=R_{\alpha n}(\tilde{a},-b,\tilde{s},\tilde{y})=0\;,\label{eq:susy}\end{equation} holds for the gauge potential $\tilde{a}$ and the coordinates $\tilde{s}$ and $\tilde{y}$ so that \begin{equation}\tilde{a}=-a\;,\ \ \tilde{s}=-s\;,\ \ \tilde{y}=-y\;.\label{eq:super}\end{equation}
 
As noted in the Introduction, the simplified BPS supersymmetry transformation between the NS-NS states and the bosonized NS-R states and between the R-R states and themselves, is the Weyl group action which exchanges the roots of $sl(2,{\boldsymbol{R}})_{{{{S}}}}$.
 Due to the S-triple algebra, \begin{equation}[\Sigma^1,\Sigma^2]_-=\Sigma^0\;,\ \ [\Sigma^0,\Sigma^1]_-=2\Sigma^1\;,\ \ [\Sigma^0,\Sigma^2]_-=-2\Sigma^2\;,\end{equation} the affine Weyl group action on the S-triple part of $\hat{{\mathfrak{g}}}$ is nothing but the transformation given in Eq.(\ref{eq:super}) except for parity transformations on the coordinates. We notice that the Yang-Mills theory is invariant under Eq.(\ref{eq:super}).
 
On the basis of this remark and Eq.(\ref{eq:susy}), we assume Eq.(\ref{eq:super2}) so that the total charge operator (the sum of the BRST and NO charges) is BPS supersymmetric, which still keeps the nilpotency.
\subsubsection{Kugo-Ojima Physical State Condition and Its Solutions}
From the Yang-Mills theory of Eq. (\ref{eq:YM}), the total currents $Q_{(n)}$ (i.e., the total charges for the time variables that are the coordinates $s_n$) for $n\in{\boldsymbol{Z}}$ of the BRST and NO transformations $\delta^{(1)}$ and $\delta^{(2)}$ are
\begin{equation}{{Q}}_{(n)}= {{Q}}_{(n)}^{(1)}+{{Q}}_{(n)}^{(2)}\;,\ \ \epsilon {{Q}}_{(n)}^{(i)}= \sum_\phi\delta^{(i)} \phi \frac{\partial {\cal{L}}}{\partial \partial_n \phi}\;,\ \ i=1,2\;.\end{equation}
Applying Eq.(\ref{eq:super2}) to all of the terms and Eq.(\ref{eq:cgauge}) to the first term, we have
\begin{equation}
{{Q}}_{(n)}= \sum_{i,l}g[a_n,b]^{i,l}\Bigl(\gamma+\frac{1}{2}ig[b,b]\Bigr)^{i,l}+Q_{(n)}^z\;.\label{eq:current}\end{equation}
Remarkably, each charge $Q_{(n)}$ is invariant under the BRST and NO transformations.\cite{KO1,KO2} Since, to show the gauge invariance, we can choose any of these charges in the Kugo-Ojima physical state condition, by taking the certain canonical quantization procedure, we consider the charge $Q_{(0)}$ as the total S-duality charge operator $Q$, which is reduced to
\begin{equation}{{Q}}=\sum_{i,l}[a_0,b]^{i,l}\Bigl(\gamma+\frac{1}{2}ig[b,b]\Bigr)^{i,l}+Q^z\;.\label{eq:Q}\end{equation}

From Eq.(\ref{eq:relation}), it immediately follows that, for the $\hat{{\mathfrak{g}}}$ generators $\Sigma$, the operator basis of $\widehat{[a_0,b]}$ after the gauge fixing is
 \begin{equation}
 ({\hat{O}}_{{[a_0,b]}}^{\Sigma})_{{\mbox{g.f.}}}=\Pi_{[a_0,b]}^{\Sigma}\;.\label{eq:trick}
 \end{equation}We denote the vector of the coefficients of $\widehat{[a_0,b]}$ by $\tilde{c}$.

According to the canonical quantization, in the canonically conjugate part $\Theta^{i,l}$ of the charge $Q$, we replace the canonical field operators $\hat{\phi}$ by their canonical conjugates $\hat{\phi}^\dagger$. We denote the canonical part of the charge $Q$ by $Q^{i,l}$.
The Kugo-Ojima physical state condition\cite{KO1,KO2} on the vacuum $\Psi[\hat{g},s]$ of the modulus parameter $\hat{g}$ and the infinite number of time variables that are the coordinates $s$ is
\begin{equation}\Biggl(\sum_{i,l}\Theta^{i,l}\biggl(\partial_n{{a}}_n+\frac{1}{2}i{{g}}[{{b}},{{b}}]\biggr)^{i,{{l}}}+\zeta\Biggr)\Psi[\hat{g},s]=0\;,\label{eq:brst}\end{equation}
where the last term $\zeta$ is proportional to the center $z$.

The solutions of Eq.(\ref{eq:brst}) in the Kac-Peterson form of the theta function are\cite{KP}
\begin{equation}\Psi(\hat{g})=\sum_{n\in{\boldsymbol{Z}}_{\ge0}}\sum_{l=0}^{n-1}\chi^{n,l}\Psi_{n,l}(\hat{g})\;,\ \ \Psi_{n,l}(\hat{g})=
\sum_{p\in{\boldsymbol{Z}}}
\Delta_{pn}O(Y)(Q^{0,l})(\hat{g})|v\rangle\;,\label{eq:sol}\end{equation}
with $\Psi_{0,l}=|v\rangle$ for the highest weight vector $|v\rangle$ of the ${{U}}(\hat{{\mathfrak{g}}})$ representation $(\varrho,|v\rangle)$, a parameter $\hat{g}$, operators $\Delta_n=(\partial/\partial s_{-n})\cdot ad_\Pi(J_n)$ ($n\neq0$) and $\Delta_0=id$ which act on field operators Eq.(\ref{eq:gauge}), operator ${{O}}(Y)$ that will be defined later, and $c$-numbered coefficients $\chi^{n,l}$. We denote the operator part in $\Psi(\hat{g})$ which acts on $|v\rangle$ by $\varrho_0(\hat{g})$.

 Each solution $\Psi_{n,l}$ with $n\ge2$ in Eq.(\ref{eq:sol}) has spontaneously broken symmetries. We denote the unbroken BRST charge by ${Q}_{br}$.
As will be explained shortly, the spontaneous symmetry breakdown is caused by the non-zero term $\zeta$ in Eq.(\ref{eq:brst}).

To complete the definition of Eq.(\ref{eq:sol}), we introduce the charge operators $Q_n$ and their dual variables $Y^n$ by (in the following, $\{,\}$ represents the superbracket)
 \begin{equation}
 Q_n=Q|_{d{{y}}^{n}}\;,\ \ \{Q_m,{Y}^n\}=\delta_{mn}\;,\ \ m,n\in{\boldsymbol{Z}}\;,
 \end{equation}
 which satisfy an infinite number of relations
 \begin{equation}
\{Q_m,Q_n\}=0\;,\ \ \{Y^m,Y^n\}=0\;,\ \ \{Y^l,[\Delta_m,Q_n]_-\}=0\;,\ \ \{Q_l,[\Delta_m,Y^n]_-\}=0\;.
 \end{equation}
 In Eq.(\ref{eq:sol}), the operator ${{O}}(Y)$ is defined by
 \begin{eqnarray}
  {{O}}(Y)=\exp\Biggl(\sum_{n\in{\boldsymbol{Z}}} \alpha_n{Y}^{n} \Delta_n\Biggr)\;,\label{eq:expO}
 \end{eqnarray}
with certain constants $\alpha_n$. Eq.(\ref{eq:expO}) is consistent, since in the superalgebra description, the operators $Y^{n}\Delta_n$ for $n\in{\boldsymbol{Z}}$ generate a nilpotent algebra.
 They satisfy an infinite number of relations
\begin{equation}\{Q_n,{{O}}(Y)\}=\epsilon\alpha_n\Delta_n{{O}}(Y)\;,\ \ \epsilon^2=0\;.
\label{eq:alpha}\end{equation}
Due to Eq.(\ref{eq:alpha}), we obtain \begin{eqnarray}{Q_n}\Delta_{pn}O(Y)(Q^{0,l})|v\rangle=(\epsilon\alpha_n\Delta_{(p+1)n}O(Y)(Q^{0,l})-\Delta_{pn}O(Y)(Q^{0,l})Q_n)|v\rangle\;.\end{eqnarray}
$Q_n|v\rangle$ is proportional to $|v\rangle$ with a ratio equal to the zero point energy which is zero due to the world sheet supersymmetry (i.e., the degeneracy between the $\Sigma^1(l)$ state and the $\Sigma^2(l)$ state for $l\in{\boldsymbol{Z}}$). This is\begin{equation}{Q_n}|v\rangle=0\;.\end{equation}
 The solution $\Psi_{n,l}$, which satisfies \begin{equation}\sum_{p\in{\boldsymbol{Z}}}\alpha_{pn}=0\;,\end{equation} with $n\ge0$ can be checked by acting with ${{{Q}_{br}}}$ on Eq.(\ref{eq:sol}).
\subsubsection{Properties of the Solutions}
Since the logarithm of each operator $\Delta_{pn}O(Y)$ in Eq.(\ref{eq:sol}) transforms one BRST transformed state $Q^{0,l}(\hat{g})|v\rangle$ into another, using the Poisson sum formula for each wave function $\Psi_{n,l}$ and its Fourier dual defined on $\phi^l$-space, in which the wave function contracts with the Fourier kernel by the field configuration $\phi$ and the index $l$ of the Heisenberg algebra, and on its conjugate space respectively the discrete modular symmetry of the wave function via the linear fractional action on the coupling constant and the field operators \begin{equation}\gamma\hat{g}= \frac{a\hat{g}+b}{c\hat{g}+d}\;,\ \ \gamma\hat{\phi}= \frac{\hat{\phi}}{c\hat{g}+d}\;,\ \ \gamma=\left(\begin{array}{cc}a&b\\c&d\end{array}\right)\;,\end{equation} follows:
\begin{equation}(\Psi_{\gamma \hat{\phi}})_{n,l}(\gamma \hat{g})=\sum_{k=0}^{n-1}u_{k,l}(\gamma)(\Psi_{\hat{\phi}})_{n,k}(\hat{g})\;,\ \ \gamma\in \Gamma\;,\label{eq:T-dual}\end{equation}
 with a unitary factor $u_{k,l}(\gamma)$ for the $l$ part of the basis $\Psi_{n,l}$ of the solutions in Eq.(\ref{eq:sol}). In the case $\gamma=\left(\begin{array}{cc}0&-1\\1&0\end{array}\right),$ Eq.(\ref{eq:T-dual}) is, for $n\in{\boldsymbol{N}}$,
\begin{equation}
(\Psi_{\frac{1}{\hat{g}}\hat{\phi}})_{n,l}\biggl(-\frac{1}{\hat{g}}\biggr)=\sqrt{-i\hat{g}n}\sum_{k=0}^{n-1}\exp\biggl(-\frac{2\pi ikl}{\hat{g}n}\biggr)(\Psi_{\hat{\phi}})_{n,k}(\hat{g})\;.\label{eq:modular}
\end{equation}

From this mathematical evidence, in the weak string coupling region (if necessary by taking S-duality), we interpret the coupling constant $g$ as the modulus parameter of F-string world sheets. In the following, we exclude the exceptional solutions which violate this discrete modular symmetry from consideration.

In the presence of the term $\zeta$ in Eq.(\ref{eq:brst}), the affinized symmetries are broken spontaneously by the generators whose central extension parts are indexed by ${\boldsymbol{Z}}_{{{\boldsymbol{N}}}}=\bigoplus_N{\boldsymbol{Z}}_N$ with (the integrals over the momenta are suppressed and the momentum conditions hold)
\begin{equation}\sum_{i=0}^2\sum_{l_M=0}^{N_M-1}\cdots\sum_{l_1=0}^{N_1-1}\Biggl(\prod_{k=1}^M((\tilde{c}\Pi_{[a_0,b]})^\dagger \kappa c_0\Pi_{a_0})^{i,l_k}\Biggr)\Psi=-\zeta\Psi\;,\label{eq:zeta}
\end{equation}
where we take the low energy limit of $g$ to $0$ and we put $\dim {\boldsymbol{N}}=M$. For the unbroken parts, the quantity corresponding to the l.h.s. of Eq.(\ref{eq:zeta}) is zero.
Eq.(\ref{eq:zeta}) means that the r.h.s., which is proportional to the number of the strings with all of multiplicities possessed by $\Psi$, is equal to the l.h.s., which is the number of the $l$-multiple strings where $l$ belongs to ${\boldsymbol{Z}}_{{\boldsymbol{N}}}$.
Thus the representation on $\Psi$ satisfies that in Eq.(\ref{eq:zeta}) the elements of the universal enveloping algebra, which is generated by the generators of the ${\boldsymbol{Z}}_{{\boldsymbol{N}}}$ part of the affinized symmetry, multiplicatively generate the non-zero weight states of the universal enveloping algebra of the affine Lie algebra $\hat{{\mathfrak{g}}}$.
The physical meaning of this statement is that, owing to the existence of the Chan-Paton charges, {{the low energy effective theory of M-theory vacua is the Yang-Mills theory with the gauge group $\bigoplus_N U(N-1)$ of the Chan-Paton fields on the multiple D-strings with ${\boldsymbol{N-1}}$ multiplicities.}}
\section{{Probe of the Space-time}}
\subsection{Nine-Spatial Degrees of Freedom}
 Our model differs from type IIB string theory on the one point that our model does not contain the graviton modes of NS-NS F-strings. In this section, we define the space-time d.o.f. by the probe of them by F- and D-strings with their modulus parameter $\hat{g}$ and the coordinates $s$.

We start by pointing out a mathematical fact in our model. Due to the Jacobi identity between the Neveu-Schwarz partition function and the Ramond partition function, the results of the Gliozzi, Scherk and Olive (GSO) projection in superstring theory\cite{Modular1,Modular2} allow us to interpret the number of cusps $\lambda$ of $\Gamma\backslash({\boldsymbol{Q}}\cup i\infty)$, where $\Gamma$ acts on $\Psi_{\hat{\phi}}(\hat{g})$ (i.e., the partition function of the doublet of Neveu-Schwarz or Ramond spectra) in Eq.(\ref{eq:T-dual}), as the number of independent physical transverse vibrations of F-strings and the degeneracy of the states at the lowest excitation level. As there are eight cusps the degeneracy of states is also eight.

In our model, we use the GSO result and expand the field operators $\hat{\phi}$ in terms of the vector $(\lambda_i)_i$ of the independent physical transverse directions as
\begin{equation}
c_\phi^{i,l}(s,p)\Pi_\phi^{i,l}(p)=\sum_{{\lambda}} (c_\phi^{i,l})_{{\lambda}}(s,p)(\Pi_\phi^{i,l})_{{\lambda}}(p)\;.\label{eq:pol}
\end{equation}
Here, the two operators $(\Pi_\phi^{i,l})_{\lambda}(p)$ and $(\Pi_\phi^{i^\prime,l^\prime})_{\lambda^\prime}(p^\prime)$ satisfy the algebra of $\Pi_\phi^{i,l}(p)$ and $\Pi_\phi^{i^\prime, l^\prime}(p^\prime)$ apart from the overall factor $\delta_{\lambda,\lambda^\prime}$.

We define two kinds of components of $\Psi$ with the field operators $\hat{\phi}_{\lambda}$ restricted towards the transverse direction ${\lambda}$ according to the decomposition in Eq.(\ref{eq:pol}) by
\begin{equation} {{{x}}}_i\Psi=\varrho_0|_{\hat{\phi}_{\lambda_i}}|v\rangle\;,\ \ {{x}}_{9}\Psi=\Psi\;.\label{eq:st}
\end{equation}

We note that the operator $\varrho_0$ on $|v\rangle$ is the multiplication of the eight operators $\varrho_0|_{\hat{\phi}_{\lambda}}$. As a result, the elements in Eq.(\ref{eq:st}) are orthogonal to each other.

We denote the vector of indices of these components of $\Psi$ by\begin{equation}(\hat{\lambda}_{\hat{i}})_{\hat{i}}=((\lambda_i)_i,\lambda_9)\;.\label{eq:angle}\end{equation}

We define the $2$-state product structure $M_2^Q$ in the hierarchy of the homotopy associative ($A_\infty$) algebra started from the BRST charge $M_1^Q=Q$ by\cite{N}
\begin{eqnarray}M_2^Q(\Psi_1,\Psi_2)=(-)^{\epsilon(\Psi_1)}\langle  a,1,2|S_{ab}\rangle|\Psi_1\rangle|\Psi_2\rangle\;.\label{eq:com}
\end{eqnarray} The $3$-vertex states $|1,2,3\rangle$ of the D-string interactions for $|p\rangle=e^{ix p}|v\rangle$ with the variables of the representation space $x_n$ are, for example, (of course due to the momentum conditions on field operators, we need not restrict subjects of the probe to be D-strings)
\begin{eqnarray}|1,2,3\rangle=&&\exp\Biggl(\sum_\phi\sum_{a,b}\sum_{l\in{\boldsymbol{N}}}\sum_\lambda\frac{(-)^{l+1}}{l}(\hat{\phi}_\lambda^{0,-l})^a(\hat{\phi}_\lambda^{0,-l})^b\Biggr)\nonumber\\&&\int \prod_{a} dp^a\delta\biggl(\sum_{a}p^a\biggr)\bigotimes_{a}|{p^a}\rangle\;,\label{eq:M1}
\end{eqnarray} which has the cyclic symmetry on the ordering of vertexes. The inverse reflector of propagating D-strings $|S_{ab}\rangle$, which satisfies the properties $O^{a}|S_{ab}\rangle=(O^T)^{b}|S_{ab}\rangle$ ($O$ is arbitrary operator) and $\langle p^a|S_{ab}\rangle=|p^b\rangle$\cite{N}, is
\begin{eqnarray}
|S_{ab}\rangle=\exp\Biggl(\sum_\phi\sum_{l\in{\boldsymbol{N}}}\sum_\lambda \frac{(-)^{l+1}}{l}(\hat{\phi}_\lambda^{0,-l})^{a}(\hat{\phi}_\lambda^{0,-l})^{b}\Biggr)
\int dp^{a}dp^{b}|{p^{a}}\rangle\otimes |{-p^{b}}\rangle\;.\label{eq:M2}
\end{eqnarray}

For the Hilbert space $V$ of vacua in Eq.(\ref{eq:brst}) and the multi-linear product structures $M_n^Q$ from $V^{\otimes n}$ to $V$ $(n=1,2,\ldots)$, the $A_\infty$ relations consist of
\begin{subequations}
\begin{align}
&\sum_{k=1}^n\sum_{i=1}^{n-k+1}(-)^\ast M_{n-k+1}^Q(\Psi_1,\ldots,M_k^Q(\Psi_{i},\ldots,\Psi_{i+k-1}),\ldots,\Psi_n)=0\;,\label{eq:com2}\\
&\ast =\epsilon(\Psi_1)+\cdots +\epsilon(\Psi_{i-1})\;.
\end{align}
\end{subequations}
To maintain the $A_\infty$ relations on the product structures, we need the sign factors $(-)^{\epsilon(\Psi)}$ in Eqs. (\ref{eq:com}) and (\ref{eq:com2}). The integer $\epsilon$ is the $(-1)$-shifted ghost number. The ghost numbers are logarithmically assigned on the bases of the ${{U}}(\hat{{\mathfrak{g}}})$-weight module with the Grassmann numbers in Eq.(\ref{eq:gauge}). Since the ghost is a Majorana particle in our model, the operators to create $l$ ghosts or $l$ anti-ghosts have $l$ ghost number for $l\in{\boldsymbol{Z}}$, and the gauge potential has zero ghost number.

The components of a degree $2$ product structure of the transverse d.o.f. $(\hat{{{x}}}_{\hat{i}})_{\hat{i}}$, which linearly act on the ${{U}}(\hat{{\mathfrak{g}}})$-module, are interpreted as the string-probe with the parameters $\hat{g}$ and $s$ of the Utiyama gravitational gauge field $\Gamma$\cite{U}
\begin{equation}
\Gamma_{\hat{i}\hat{j}}^{\hat{k}}=\langle M^Q_2(\hat{{{x}}}_{\hat{i}}\Psi,\hat{{{x}}}_{\hat{j}}\Psi),\hat{{{x}}}_{\hat{k}}\Psi\rangle\;,\label{eq:gravity1}
\end{equation}
which will be restricted to a real number value in the next paragraph. 
This is because Eq.(\ref{eq:gravity1}) is the gauge field of the infinitesimal (i.e., linear) deformations of the vector $(\hat{{{x}}}_{\hat{i}}\Psi)_{\hat{i}}$ by the vector $(\hat{{{x}}}_{\hat{i}}\Psi)_{\hat{i}}$ conserving the form of the string interactions in the BRST string field theory.\cite{HIKKO} We note that the Utiyama field was introduced as the gauge field of the local affine transformations (the general coordinate transformations) on the space-time coordinates, and the BRST string field theory contains such the gauge symmetry in the low energy limit. In our model, we consider such the field on the nine spatial coordinates.

{{The spatial expanses of the probes along the section $\Psi$ exist in the fiber directions of the principle bundle.}} In order to describe them, we introduce the real number valued time variables $\hat{x}_{\Sigma,\hat{k}}$ of the generators $\Sigma_{\hat{\lambda}_{\hat{k}}}$ in the fiber space and define the spatial coordinates of the wave function with the variables by $\hat{x}_{\Sigma,\hat{k}}(\hat{x}_{\hat{k}}\Psi[\hat{g},s])$. The spatial expanses are generated by the exponential maps of the $\Gamma$-fields with the infinitesimals $d\hat{x}_{\Sigma,\hat{k}}$ of the wave function and the generators corresponding to the spatial parts of the general coordinate transformations of the {\it{space}} labeled by the time variables $(\hat{x}_{\Sigma,{\hat{k}}})_{\Sigma,\hat{k}}$. The time variable $x_{\Sigma^i(l),\hat{k}}$ represents the $\hat{k}$-th spatial coordinate probed by $l$ strings of the $i$-th kind (i.e., R-R, NS-NS, or NS-R).

If two cusps $\lambda_a$ and $\lambda_b$ belong to the same $\Gamma$-class of cusps, 
\begin{equation}
{{x}}_{a}(\Psi_{\gamma\hat{\phi}})_{n,l} =\sum_{k=0}^{n-1} u_{k,l}(\gamma){{x}}_{b} (\Psi_{\hat{\phi}})_{n,k}\;,\ \ \gamma \lambda_a=\lambda_b\;,
\end{equation}
holds and the action $\gamma$ on the cusp preserves the weight of the state $x_a\Psi_{\gamma\hat{\phi}}$. Thus the product structure $M_2^Q$ and its components are retained.
The physical meaning of this statement is that {{the dimensionality of the probed eight space d.o.f. is determined by the symmetry group $\Gamma$ in Eq.(\ref{eq:T-dual}).}}

 {{The spatial expanses of the space-time $\hat{x}_{\hat{i}}$ for $\hat{i}=1,2,\ldots,9$ depend on how we choose to probe them. Their possible descriptions produce the same physics of correlators via their wave functions equivalent to each other such as in the mirror symmetry phenomena.\cite{Mirror1,Mirror2,Mirror3} The probe can be generalized to an analyzing wavelet. Eq.(\ref{eq:com}) is recognized as a continuous wavelet transformation of the wave functions. Thus, the wavelet analysis of the wave functions would systematically generalize the geometry of the target spaces.}}

\subsection{Cosmic Time}
The cosmic time $\tau$ is the affine parameter assigned on spatial hypersurfaces sliced from space-time. As will be seen in the next section, our formulation of type IIB string theory vacua is close to that of the wave functions of the Universe. Then, to quantize the Universe, the treatment of the cosmic time needs to be trivial, in other words, the physical quantities do not depend on the choice of the cosmic time. Actually, to quantize the Universe, we decompose the space-time metric by the Arnowitt-Deser-Misner formalism\cite{ADM} and after the variation of the action by the Lapse function we set the Lapse function at unit. So, our wave function has no information about the potential of the increment of the cosmic time that is the temporal-temporal part of the gravitational potential (i.e., the space-time metric) and our definition of the increment of the cosmic time needs to be done not by the space-time metric but by a clock of the string excitations.
{{We define the increment of the {{cosmic time}} $\delta{\tau}(s)$, to describe the change of the system only (see Section 4.2), in units of the Planck time as the operator, whose expectation value in the system is for the expectation values of the momenta ${p}^0$

\begin{equation}\la\delta{\tau}(s)\ra=\biggl[\frac{ k\la Q\ra}{\la \sqrt{(\Omega_0-\Omega)({p}^0)}\ra}\biggr](s)\;,\label{eq:time1}\end{equation}where the numerator of Eq.(\ref{eq:time1}), $k{Q}(s)$, is proportional to the Hamiltonian of the system ${Q}(s)$ 
 as the frequency times the number of the elements of the system (when we consider the free part of it),
and the denominator of Eq.(\ref{eq:time1}) is defined by the square root of the shifted minus Casimir operator $\Omega$ of the representation $\Pi(p)$ (see Eq.(\ref{eq:gauge})) of the affine Lie algebra $\hat{{\mathfrak{g}}}$,\cite{Casimir,VGKAC2}
where we invoke the identity for an arbitrary functional $f$ of ${\Omega}$
\begin{eqnarray}
\la w|f({\Omega})\delta \tau|w\ra&&=\sum_{w^\prime}\la w|f({\Omega})|w^\prime\ra \la w^\prime|\delta \tau|w\ra\nonumber\\
&&=\la w|f({\Omega})|w\ra\la w|\delta \tau|w\ra\;.
\end{eqnarray}
In Eq.(\ref{eq:time1}) $\Omega_0$ is the maximum value of the Casimir operator.

The coordinates $s$ have no relevance to the history of the cosmic time $\tau$ but only its increment $\delta\tau(s)$.
The solutions $\Psi$ of Eq.(\ref{eq:brst}) do not have the variable of the history of the cosmic time $\tau$, so we add it to $\Psi$.

These definitions are equivalent to the equation
\begin{equation}\sqrt{ (\Omega_0-\Omega)({p}^0)}\delta {\tau}(s)= k{Q}(s)\;.\label{eq:grav1}
\end{equation}
We notice that the Casimir part of Eq.(\ref{eq:time1}) has dimensions of time, since $\hbar$ times the square root of the shifted minus Casimir of the representation $\Pi(p)$ (that is, the sum of $\hbar$ times the time frequencies of string excitations $\omega_{p^0}$ in the field operators of Eqs.(\ref{eq:gauge})) has dimensions of energy. The string excitations and the time periods corresponding to these frequencies $\omega_{p^0}$ are recognized as the clock and the Casimir part of  Eq.(\ref{eq:time1}) respectively.
Thus, the free part of Eq. (\ref{eq:time1}) is proportional to the quotient of the expectation value of the Hamiltonian divided by the time frequency of the string excitation $\omega_{p^0}$, that is, the expectation value of the number of the elements of the system.

The S-duality is a non-compact symmetry. So, we need to specify the kind of the S-duality part of the representation $\Pi(p)$. The unitary irreducible representations of $SL(2,{\bs{R}})$ are classified into three kinds: the principal series, the discrete series and the supplementary series.\cite{Rep} The principal series has imaginary and continuous parameters, however it results the positive Casimir (then, $\sqrt{(\Omega_0-\Omega)({p}^0)}$ and $Q(s)$ are not Hermitian).
 On the other hand, the discrete series has natural number valued highest or lowest weights. It is well known that the principal series with the discrete series provides a complete set of basis functions on ${\mathfrak{H}}$. When we require the Hermiticity of $Q(s)$, we consider the discrete series only.

The cosmic time makes sense only in terms of its change in the Schr$\ddot{{\rm{o}}}$dinger equations (see Eqs.(\ref{eq:Sch})). The issue of simultaneity will be resolved in the derived categorical formulation as the quasi-equivalence of $Q$-complexes that represent cosmic time evolutions (see Section 5.1).

 With respect to the free part of the Hamiltonian, the grounds for the definition in Eq.(\ref{eq:time1}) is as follows. As explained above, the free part of Eq.(\ref{eq:time1}) is proportional to (the expectation value of) the number of elements of the system $\la n\ra$. When we fix the expectation values of momenta $p^0$, the statistical properties of $\la n\ra(s)$ around the coordinates $s$ are those of the eigenstates of the Hamiltonian $|n\ra$ (if we restrict the Hamiltonian to its free part) which are labeled by the numbers of elements $n$. The cosmic time increment needs to count all of the non-unitary processes. The non-unitary processes induced by the Hamiltonian are classified by the transitions between the states labeled by the numbers $\la n\ra$ via their superpositions. Thus, the variance of the number $\la n\ra(s)$ around the coordinates $s$ captures the changes in the non-unitary processes induced by the Hamiltonian. Consequently, $\la n\ra$ contains the statistical properties (i.e., mean, variance and distribution function) of $\delta\tau(s)$ around the coordinates $s$, as will be discussed in Section 4.2. Here, we have discussed Eq.(\ref{eq:time1}) using coarse-grained values of some quantities, so Eq.(\ref{eq:time1}) may not be the exact form of the cosmic time increment. On this issue, further refinement may be needed, but is beyond the scope of the present investigations.

{{The increment of the cosmic time reflects the number of d.o.f. of the system which distinguish the R-R, NS-NS and NS-R strings by non-unitary processes, as will be seen in the discussion in Section 4.2.}}}}

 The GSO result is compatible with the NSR algebra\cite{NSR} that indicates the Lorentz signature of the ten-dimensional space-time metric. So, as the local space-time metric, we assume the one whose signature is Lorentzian.
\section{Cosmic Time Developments}
\subsection{The Universe: A Conjecture}
In this section, we physically interpret the wave function, which corresponds to a type IIB superstring vacuum.
As will be seen later, it is consistent that our formulation of M-theory vacua by Eq.(\ref{eq:brst}) does not explicitly depend on changes of the cosmic time. This is the same as the formulation of the quantum behavior of the early Universe by the Wheeler-De Witt equation.\cite{WDW1,WDW2,WDW3,WDW4,WDW5} The wave function of the Universe is defined on superspace, that is, the moduli space of spatial metrics and field configurations under the moduli of spatial diffeomorphisms.

The Universe is canonically quantized by the spatial metrics $h_{\hat{i}\hat{j}}$ (with spatial indices $\hat{i},\hat{j}=1,2,\ldots,9$) on the spatial hypersurfaces sliced from space-time by values of the cosmic time and their momenta. It is described by a wave function of the nine-dimensional spatial metric $h$, the axion and dilaton ${\cal{M}}$, the spatial parts of the 2-form NS-NS and R-R potentials ${{{B}}}^{(i)}_{\hat{i}\hat{j}}$ for $i=1,2$, and the cosmological constant appearing in Eq.(\ref{eq:brst}) as $\zeta$. This wave function is the solution of the WDW equation \begin{equation}{\cal{H}}\psi[h_{\hat{i}\hat{j}},{\cal{M}},{{{B}}}^{(i)}_{\hat{i}\hat{j}}]=0\;,\label{eq:WDW}\end{equation} for the quantum mechanical Hamiltonian operator ${\cal{H}}$ of type IIB supergravity as the low energy effective field theory of type IIB string theory. The spatial metric $h_{\hat{i}\hat{j}}$ is related to the Utiyama field $\Gamma_{\hat{i}\hat{j}}^{\hat{k}}$ by
\begin{equation}
\Gamma_{\hat{i}\hat{j}}^{\hat{k}}=\frac{1}{2}h^{\hat{k}\hat{l}}(\partial_{\hat{j}} h_{\hat{l} \hat{i}}+\partial_{\hat{i}}h_{\hat{j}\hat{l}}-\partial_{\hat{l}}h_{\hat{i}\hat{j}})\;,\ \ h^{\hat{i}\hat{j}}h_{\hat{j}\hat{k}}=\delta^{\hat{i}}_{\hat{k}}\;,\label{eq:Utiyama}
\end{equation}
where the indices $\hat{i},\hat{j}$ and $\hat{k}$ include the transverse d.o.f. $x_9$.
 The spatial metric $h_{\hat{i}\hat{j}}$ can be reconstructed from the Utiyama field $\Gamma_{\hat{i}\hat{j}}^{\hat{k}}$ in Eq.(\ref{eq:Utiyama}) except for the ambiguity of the local frame field on the Universe used to define the connection 1-form.
 
 As mentioned in the Introduction, the Hilbert space of the fields of gauged S-duality on the moduli space of vacua contains all of the excitations of strings except for gravitons and space-time is described by a probe using these excitations. Therefore, the consistency of our model requires that the wave function of the Universe is equivalent to the wave function defined by Eq.(\ref{eq:brst}) in its local field description\begin{equation}\psi=\Psi\;.\label{eq:Universe}
\end{equation}

{{This conjecture Eq.(\ref{eq:Universe}) states that the dynamical d.o.f. of type IIB string theory appearing in the field and metric configurations are reducible on an integrable hierarchy on the Riemann surface of $\hat{g}$ with an infinite number of deformation variables $s$. This is a novel kind of reduction of string theory. The reason why this reduction works is essentially same as the case of type IIB matrix model.\cite{IKKT} To show this reduction, we embed the l.h.s. of Eq.(\ref{eq:Universe}) in the moduli space of vacua. This process is possible according to the dictionary invented in the previous sections. In the moduli space of vacua, the affinized S-duality is gauged, and it is recognized as a set of an infinite number of independent internal local symmetries. As explained in the Introduction, they are the Chan-Paton gauge symmetries on (anti-)D-strings. Then, the Eguchi-Kawai large $N$ reduction works.\cite{EK} The dynamical d.o.f. of the gauge fields on the base space of the moduli space of vacua, which are possessed by the l.h.s. of Eq.(\ref{eq:Universe}), are reduced. We recall that the spatial d.o.f. and all of the excitations of strings are embedded in the fiber space. The reduced wave function of the Universe is described by the wave function $\Psi[\hat{g},s]$ which has no variable of the fiber space and is not explicitly dynamical.}}

We note that although the r.h.s. of Eq.(\ref{eq:Universe}) is based on the $\Gamma$-field which is not a tensor quantity with respect to the spatial part of the general coordinate transformations, the l.h.s. is based on the spatial metric of the Universe which is a tensor quantity and its physics does not depend on choices of the system of the coordinates on the Universe.

The reader may find this consequence of our conjecture to be strange, since although the wave function $\Psi$ does not depend on the cosmic time, the Universe which is the solution of the equations of motion of type IIB string action depends non-trivially on the development of the cosmic time. This paradox is resolved by the following argument. In our model, the spatial part of the space-time structure is defined via the probes of it by strings, and in Eq.(\ref{eq:com}) the wave function $\Psi$ itself does not depend on the cosmic time, but the product structure $M_2^Q$ and its higher order corrections $M_n^Q$ for $n\ge3$, which describe the probes, depend on the cosmic time. Consequently, Eq.(\ref{eq:com}) depends non-trivially on the cosmic time.

 In the l.h.s. of Eq.(\ref{eq:Universe}), the cosmic time and the dependence on it of the matter fields have been introduced for the semiclassical phase of the wave function of the Universe by substituting the definition of the momentum of the Universe (in the minisuperspace model, the time rate of the dynamical changes of the scale factor of the Universe) into the Hamilton-Jacobi equations of the matter fields obtained from the WDW equation Eq.(\ref{eq:WDW}). As a result, we obtain the Schr${\ddot{{\rm{o}}}}$dinger equations of the matter fields. In contrast, our way of introducing the cosmic time is alike the way of introducing the entropy in statistical mechanics which counts the distinguishable partitions.

\subsection{Matter and Hamiltonians}
A matter system of microscopic or macroscopically coherent quantum fields is described by the wave function ${\psi}_\Lambda$, depending on the order parameters $\Lambda$, which describe the symmetry and its breakdown in the effective vacuum.

 The uniqueness of the bare Hamiltonian of our model requires that the state ${\psi}_\Lambda$ of any such system takes the form\begin{equation}
{\psi}_\Lambda=R\psi_{\hat{v}}\;,\ \ \psi_{\hat{v}}=\varrho(\hat{v})\Psi\;,\ \ \hat{v}\in V_s\;,\label{eq:brain}
\end{equation} where $V_s$ is the state space of the system and $R$ is a renormalization transformation on the vacuum $\Psi$, in which $\psi_{\Lambda}$ keeps the d.o.f. of the symmetry of the system and the d.o.f. of the Heisenberg symmetries in Eq.(\ref{eq:affine}).

The BRST transformation is
\begin{equation}
\delta \psi_{\Lambda}=Q\psi_{\Lambda}\;.
\end{equation}
We identify the cosmic time with the time variable of the BRST charge
\begin{equation}
\delta_{\tau_R}\psi_{\Lambda}=\delta \psi_{\Lambda}\;,\ \ [\delta_{\tau_R},\tau_R]_-=i\hbar\;,
\end{equation}
where $\tau_R$ is an effectively scaled cosmic time of the renormalized matter wave functions $\psi_\Lambda$.

{{The equation of the matter wave function $\psi_\Lambda$ is\begin{equation}i\hbar\frac{\delta\psi_\Lambda}{\delta \tau_R}={{Q}}{\psi_\Lambda}\;.\label{eq:Sch}\end{equation}
The increment of the cosmic time is an operator, thus it makes a rigorous sense only for its eigen wave functions. In this paper, we consider the time developments of its eigen wave functions only, otherwise we take its expectation value in the system.

The functional form of the increment of the cosmic time $\delta\tau_R(s)$ is already determined by Eq.(\ref{eq:time1}). Eqs. (\ref{eq:Sch}) specify the functional variation $\delta/\delta \tau_R$. We note that $\delta/\delta\tau_R=\delta/\delta(\delta\tau_R)$. The time coordinates $s$ are infinitely many, matching the number of degrees of freedom of the model. This variation $\delta\psi_\Lambda/\delta \tau_R$ is between functions of the coordinates $s$. Without Eqs.(\ref{eq:Sch}), when $\tau_R$ changes its value, we do not know which of the coordinates $s$ has changed to cause this shift; Eqs.(\ref{eq:Sch}) specify it. The cosmic time is one degree of freedom among an infinite number of coordinates. If we fix the cosmic time, the wave function is a function of the remaining degrees of freedom. The variation of the wave function with respect to the cosmic time is determined by Eqs. (\ref{eq:Sch}).

To know the exact form of the functional variation $\delta/\delta \tau_R$, we need Eqs.(\ref{eq:Sch}) for the full Hilbert space. Thus the description of $\delta\tau_R(s)$ is stochastic. }}
{{Namely, the unpredictability caused by the infiniteness of the number of d.o.f. introduces the stochastic processes. This logic can be applied not only to the renormalized wave function $\psi_\Lambda$ but also to the probe of the space-time in Eq.(\ref{eq:com}), which is closed under the cosmic time processes, and on which we rely for an imperfect and partial description.}}

Due to Eq.(\ref{eq:brst}), it is consistent that the wave function $\Psi$ does not depend on changes of the cosmic time
\begin{equation}{\delta_{\tau} \Psi}=0\;.\label{eq:SchUniverse}\end{equation} 
The Hamiltonian operator ${{\cal{H}}}$ of the state space of the system $V_s$ takes the form
\begin{equation}
{\cal{H}}=(QR)|_{V_s}\;.
\end{equation}
Using this Hamiltonian, we rewrite the variation by the cosmic time in Eq.(\ref{eq:Sch}) as an average over the coordinates on the moduli space of vacua:
\begin{equation}
\langle\psi_{\Lambda}(\tau_R)\rangle\approx\exp\biggl(-\frac{i\tau_R}{\hbar} {\cal{H}} -\frac{\sigma_R\tau_R}{2\hbar^2}{\cal{H}}^2\biggr)\langle\psi_{\Lambda}(0)\rangle\;,\label{eq:est}
\end{equation}
where the average is defined by the following recursion equation
\begin{subequations}
\begin{align}
\langle \psi_{\Lambda}(\mu_R)\rangle&=\int {\cal{D}}\tau_R^\prime(s) \exp\biggl(-\frac{i\delta\tau_R^\prime(s)}{\hbar}{\cal{H}}\biggr)\langle\psi_{\Lambda}(0)\rangle\label{eq:ave}\\
&\approx\int d\tau_R^\prime\exp\biggl(-\frac{i\delta\tau_R^\prime}{\hbar}{\cal{H}}\biggr)f(\delta\tau_R^\prime)\langle\psi_{\Lambda}(0)\rangle\;.\label{eq:ave2}
\end{align}
\end{subequations}
In Eq.(\ref{eq:ave2}) we rewrite the functional integral with respect to $\delta\tau_R^\prime(s)$ as an average over a normal stochastic variable $\delta\tau_R$ with variance $\sigma_R$, mean $\mu_R$ and distribution function $f(\delta\tau_R^\prime)$.
We note that, since the nilpotency of the BRST charge requires the affinized symmetry, in general, ${\cal{H}}^2$ is not identically zero due to its restriction.

The Hamiltonian ${\cal{H}}$ is a Hermitian operator. Thus for the eigenvalues $\{\lambda\}$ of ${\cal{H}}$, there exists a unique spectral family $\{d{\cal{H}}(\lambda)\}$, and the spectral decomposition is\begin{equation}
{\cal{H}}=\int \lambda d{\cal{H}}(\lambda)\;.\label{eq:Sch2}
\end{equation}

From the elementary property of the spectral components ${\cal{H}}(\lambda)$ in Eq.(\ref{eq:Sch2}),
\begin{equation}
{\cal{H}}({\lambda_1}){\cal{H}}({\lambda_2})=\delta_{\lambda_1,\lambda_2}{\cal{H}}({\lambda_1})\;,
\end{equation}
it follows that,
\begin{equation}
{\cal{H}}^2=\int \lambda^2d{\cal{H}}(\lambda)\;,
\end{equation}
and the cosmic time development in Eq.(\ref{eq:est}) satisfies the properties of a contraction semigroup in the parameter $\tau_R$.

The d.o.f. of collapses of the superposition of wave functions \begin{equation}R\Psi=\sum_\lambda c_\lambda R\Psi^\lambda\;,\label{eq:super3}\end{equation} is the spectral component ${\cal{H}}(\lambda)$. In the superposition of Eq.(\ref{eq:super3}), each component $R\Psi^\lambda$ is distinguished from the others by the spectral components ${\cal{H}}(\lambda)$ such that \begin{equation}{\mbox{if}}\ \ {\cal{H}}(\lambda)\psi^{\lambda_1}_{\hat{v}}\neq0\;,\ \ {\mbox{then}}\ \ {\cal{H}}(\lambda)\psi^{\lambda_2}_{\hat{v}}=0\;,\end{equation} for $\lambda_1\neq \lambda_2$ and elements $\hat{v}$ of the state space ${V}_s$ of the system. Concretely, the spectral component ${\cal{H}}(\lambda)$ is defined by the restriction of ${\cal{H}}$ on the part which lies within the eigenspace $V_\lambda$ for eigenvalue $\lambda$,\begin{equation}{\cal{H}}(\lambda)={\cal{H}}|_{V_\lambda}\;,\ \ V_s=\bigoplus_\lambda V_\lambda\;,\end{equation} which induces a non-unitary action on the wave function within the non-zero variance of the increment of the effectively scaled cosmic time as an operator of the contraction semigroup in the cosmic time evolution:\begin{equation}\Delta(\tau_R)\approx\exp\biggl(-\frac{i\tau_R}{\hbar}{\cal{H}}-\frac{\sigma_R\tau_R}{2\hbar^2}{\cal{H}}^2\biggr)\;.\label{eq:timeexact}\end{equation}

We regard the estimation of $\delta\tau_R(s)$ from the incomplete knowledge of it (i.e., Eq.(\ref{eq:Sch})) as a normal stochastic process of the variable $\delta\tau_R$ via Eq.(\ref{eq:ave2}) such that the probability $P_\lambda$ of the collapse into the branch $R\Psi^\lambda$ is 
\begin{equation}
P_\lambda=\langle R\psi_{\hat{v}},{\cal{H}}(\lambda)\psi_{\hat{v}}\rangle=|c_\lambda|^2\;,\ \ \sum_\lambda P_\lambda=1\;.
\end{equation} If we know the exact form of $\delta/\delta \tau_R(s)$, all of the d.o.f. of both the non-unitary and unitary cosmic time developments of Eq.(\ref{eq:brain}) are reducible to an infinite number of coordinates $(s_n)_n$ on the moduli space of vacua.
\section{Non-perturbative Description Using the Non-linear Potential}
\subsection{Derived Category Structure Using Wave Functions}
In our modeling, due to Eqs.(\ref{eq:SchUniverse}) and (\ref{eq:est}), the perfect description of the Universe is independent of changes in the cosmic time, and non-trivial cosmic time processes can be applied only to closed systems with imperfect, partial descriptions and a non-zero retention time of the superposition of the wave functions.
\footnote{Due to Eq.(\ref{eq:est}), the retention time of the superposition of the wave functions tends to zero for the macroscopic objects.} Systems which lose the retention time of the superposition of the wave functions
 have a 
 classical cosmic time evolution and are essentially removable objects, whereas systems with a non-zero 
 retention time of the superposition of the wave functions
 genuinely constitute a quantum mechanical world with common cosmic time processes such as quantum mechanical branching. That is, for the system with the non-zero retention time of the superposition of the wave functions, the variance of the increment of the cosmic time induces the non-unitary time development of a system. By Eq.(\ref{eq:brain}) a system is a state space $RV_s$ of a $U(\hat{\mathfrak{g}})$-module $V_s$ with a certain renormalization $R$ and its time development is mapped to the projective resolution of the diagram of\begin{equation}\xymatrix{0& RV\ar[l]_Q}\end{equation} as the $Q$-complex \begin{equation}\xymatrix{0& RV\ar[l]_Q& P_1RV\ar[l]_Q&\cdots\ar[l]_Q}\label{eq:exact}\end{equation}where in the $n+2$-th element of Eq.(\ref{eq:exact}) we restrict both of $QR$ and $\Psi$ to the same state space with the fixed $n$-th cosmic time value counted by the events of non-unitary processes.

{{Since in our context, the Kugo-Ojima physical state condition means that the wave function is an eigenfunction of the Hamiltonian with zero eigenvalue, the non-unitary time process, that is, the collapse of a superposition of wave functions changes the eigenvalues of the Hamiltonian. Namely, for the eigenvalue $\lambda$ of the Hamiltonian
 \begin{equation}
 {\mb{ker}}Q=RV_{\lambda=0}\;,\ \ {\mb{im}}Q=RV_{\lambda\neq0}\;,
 \end{equation}
 holds. Then, the kernel of $Q$ does not match the full state space $V$ and the cohomology of $Q$ is non-trivial, that is, not the full state space. We note that, generally, superpositions of the wave functions are generated by unitary time processes in the larger system. Thus, a non-unitary process may occur at any cosmic time.}}

By the $Q$-cohomology content in Eq.(\ref{eq:est}) only, each system is specified and the cohomologically non-trivial content is due to the non-unitary second factor in Eq.(\ref{eq:est}). For a macroscopic physical object, we can interpret this as a collection of microscopic quantum states with non-trivial effects of time variances or as a large-scale macroscopic quantum state with trivial effect of time variance. 
These interpretations need to be unified.
These observations lead us to the derived category description of the quantum mechanical world under the moduli of quasi-isomorphism equivalences of the BRST complexes. We denote by $D(C)$ this derived category of the BRST complexes of the base abelian category $C$.
   Due to Eq.(\ref{eq:Whitham}), the quasi-isomorphisms which commute with the cosmic time development by the $Q$ operation are given by renormalizations. 
Here, the derived category $D(C)$ of a base abelian category $C$ is defined by restriction of the homotopy category $K(C)$ on a closed system of the products of quasi-isomorphisms in $K(C)$.\cite{GM}
 The objects of our base abelian category $C$ are the spaces of states created from a given wave function $\Psi$ by the actions $\varrho(V)$, i.e., the $U(\hat{\mathfrak{g}})$-weight modules. The morphisms are the transformations compatible with the differential $Q$ or the covariant derivative $\nabla$ (namely, where $Q$ or $\nabla$ is a vector) respectively. (We denote the latter base abelian category by $C^\nabla$.)
When we consider the theory of gauged S-duality using the linear wave function $\Psi$ only, the morphisms of base abelian category $C^Q$ are defined only by the homomorphisms, denoted by $k$, compatible with the differential $Q$. Then, the morphisms, denoted by $h$, between two complexes are defined by $h=kQ+Qk$.

\subsection{Non-linear Potential}
Based on this derived category structure $D(C)$ of the quantum mechanical world description, we generalize the results in the last section by a substantially different method. We introduce a single master equation as the generalization of the Kugo-Ojima physical state condition for a non-linear potential, denoted by $\aleph$ (standing for the symbol `{\it{$A_\mu$}}' of a gauge potential), which describes the non-peturbative effect or dynamics alluded in the Introduction, according to the following three guiding principles. As the concrete form of the equation, we adopt a single vanishing curvature.

\renewcommand{\theenumi}{\roman{enumi}}
\renewcommand{\labelenumi}{\theenumi)}

\begin{enumerate}
\item
{{The local principle.}} In our modeling, it is gauged S-duality.
\item The generalized gauge covariance using the non-linear potential.
\item The equation vanishes under the action of the covariant derivative, due to the generalized gauge invariance (we note that the covariant derivative is the generalization of the BRST charge $Q$).
\end{enumerate}

As the result, the generalized Kugo-Ojima physical state condition is regarding operator valued $\hat{\mathfrak{g}}$-connection $\aleph$ on the fiber space:\footnote{In the following, we denote the BRST charge and the NO charge by $Q$ and $\bar{Q}$ respectively. So, the notation $Q$ used here is different from that in Eq.(\ref{eq:Q}).}
\begin{equation}\Omega-\frac{1}{2}[[\bar{\na},\Omega],\na]=0\;,\label{eq:master}\end{equation}
where $\Omega$ is the curvature form
\begin{equation}
\Omega=[\na,\na]\;,
\end{equation} and we introduce $\bar{\aleph}$ as the dual vector field, whose coefficients $\bar{\aleph}_i$, indexed by the differential basis $\partial_i$ on the tangent bundle for the dual coordinates ${\cal{Y}}^i$ of $Q_i=Q|_{dy^i}$, are the operators made from the dual basis of $\hat{{\mathfrak{g}}}$. These coefficients are canonically conjugate to the one of $\aleph$, indexed by the 1-form basis $d{\cal{Y}}^i$ on the cotangent bundle, as \begin{equation}[{\aleph}_i,\bar{\aleph}_j]=\delta_{ij}\;,\label{eq:can}\end{equation} and $\nabla$ and $\bar{\nabla}$ are the covariant derivatives on the categorical $Q$-complexes
\begin{equation}\nabla{\cal{O}}=\delta{\cal{O}}+[\aleph,{\cal{O}}]\;,\ \ \bar{\nabla}{\cal{O}}=\bar{\delta}{\cal{O}}+[\bar{\aleph},{\cal{O}}]\;,\end{equation}for an arbitrary operator valued form ${\cal{O}}$ and $[,]$ is the ${\boldsymbol{Z}}$-graded 
commutator for an arbitrary pair of a $d_a$-form ${\cal{O}}_a$ and $d_b$-form ${\cal{O}}_b$:\footnote{The action of the differential basis on a commutator of forms satisfies a Leibniz rule similar to the one for $Q$ (see Eq.(\ref{eq:Leibniz})). It is consistent due to $d{\cal{Y}}^i\wedge d{\cal{Y}}^i=0$.}
\begin{equation}[{\cal{O}}_a,{\cal{O}}_b]={\cal{O}}_a\wedge {\cal{O}}_b-(-)^{d_ad_b}{\cal{O}}_b\wedge {\cal{O}}_a\;,\end{equation}
which satisfies the super Jacobi identity
\begin{eqnarray}
(-)^{d_ad_c}[{\cal{O}}_a,[{\cal{O}}_b,{\cal{O}}_c]]+(-)^{d_bd_c}[{\cal{O}}_c,[{\cal{O}}_a,{\cal{O}}_b]]+
(-)^{d_ad_b}[{\cal{O}}_b,[{\cal{O}}_c,{\cal{O}}_a]]
=0\;.\label{eq:Jacobi}
\end{eqnarray}
Here, we consider the super Lie algebra of $\hat{{\mathfrak{g}}}$ as introduced in Section 2, and the degree $d$ of the element ${\cal{O}}$ is its ghost number. We note that in general, ${\cal{O}}_a\wedge {\cal{O}}_b\neq-{\cal{O}}_b\wedge{\cal{O}}_a$ for $1$-forms, since we treat the product of matrices in $\hat{{\mathfrak{g}}}$ and the outer product of forms simultaneously. 
Since the BRST differential has ghost number one, the ghost number coincides with the degree of the element as a form. The space of the operators $O$ splits into $\oplus_{i\ge0} O^i$ labeled by the ghost number $i$ with $[O^i,O^j]\subset O^{i+j}$. The BRST differential shifts $O^i$ to $O^{i+1}$ and acts on commutators of operators as \begin{equation}\delta[{\cal{O}}_a,{\cal{O}}_b]=[\delta{\cal{O}}_a, {\cal{O}}_b]+(-)^{d_a}[{\cal{O}}_a, \delta{\cal{O}}_b]\;.\label{eq:Leibniz}\end{equation} 

We check the requirements of the three principles in Eq.(\ref{eq:master}).
The first principle requires that infinitesimal deformations $\Psi$ of the parallel section to $\nabla$ obey the linearized equation Eq.(\ref{eq:brst}).
The principle of covariance requires that the non-linear potential $\aleph$ obey the equation written only using $\nabla$ and $\bar{\nabla}$. Eq.(\ref{eq:master}) satisfies these requirements. Finally, to show the third principle on Eq.(\ref{eq:master}), we contract the indices $i$ and $b$ of the Bianchi identity
\begin{equation}
\nabla_i([\nabla_j,\na_k])_{ab}+\nabla_k([\nabla_i,\na_j])_{ab}+\nabla_j([\nabla_k,\na_i])_{ab}=0\;,
\end{equation}for the components of $[\nabla,\nabla]$
 obtained from the super Jacobi identity in Eq.(\ref{eq:Jacobi}). For the components of $\nabla$, we have\begin{eqnarray}
 [\nabla_i,[\nabla_j,\nabla_k]]+[\nabla_k,[\nabla_i,\nabla_j]]+ [\nabla_j,[\nabla_k,\nabla_i]]=0\;,
\end{eqnarray}where we use the fact that $\nabla$ has ghost number $1$. We 
denote the matrix elements (not in the sense of the expectation values) of the operator $[\nabla_i,\na_j]$ by $([\nabla_i,\na_j])_{ab}$, and these are defined by \begin{equation}[\nabla_i,\nabla_j]{\cal{O}}^a=([\nabla_i,\na_j])^a_{b}{\cal{O}}^b\;,\end{equation} for an arbitrary operator valued $\hat{{\mathfrak{g}}}$-connection ${\cal{O}}$ on the fiber space. The indices $a$ and $b$ denote the bases of $\hat{{\mathfrak{g}}}$, and the contraction of indices is taken using the metric on the fiber space. By analogy with the $c$-valued curvature tensor, we assume the symmetry of the indices of $[\nabla,\nabla]$
\begin{equation}
([\nabla_i,\na_j])_{ab}=([\nabla_a,\na_b])_{ij}=-([\nabla_j,\na_i])_{ab}=-([\nabla_i,\na_j])_{ba}\;.\label{eq:Riemann}
\end{equation} Then, the Bianchi identity takes the form 
\begin{eqnarray}\nabla^iG_{aijk}
&=&0\;,\\
G_{aijk}
&=&([\nabla_j,\na_k])_{ai}+\delta_{ij}\sum_i([\nabla_k,\na_i])_{ai}-\delta_{ik}\sum_i([\nabla_j,\na_i])_{ai}\;.\label{eq:G}
\end{eqnarray}
Using the Leibniz rule, the equality \begin{equation}\bar{\delta}{\aleph}+\delta\bar{\aleph}=0\;,\end{equation} and Eq.(\ref{eq:Riemann}) for the interchange of the indices $i$ and $j$ for the action of the component $\bar{\na}_i$ on the component $[\nabla_j,\na_k]$ and using the Leibniz rule and Eq.(\ref{eq:Riemann}) for the interchange of the indices $a$ and $b$ for the action of the dual basis $\bar{S}^i$ of $\bar{\na}=\bar{\na}_i\bar{S}^i$ on the basis $S^a$ and $S^b$ of $[\nabla_j,\nabla_k]=([\nabla_j,\na_k])_{ab}S^aS^b$ indexed by $a$ and $b$ such that $\bar{S}^aS^b=\delta_{ab}$ (the factor $1/2$ in the second term of Eq.(\ref{eq:master}) comes from this action), and keeping in mind the canonical conjugation relations Eq.(\ref{eq:can}), we find the first term and the sum of the second and third terms of this quantity $G$ to be locally the first and second terms of the l.h.s. of Eq.(\ref{eq:master}) respectively.
Consequently, Eq.(\ref{eq:master}) satisfies the third principle of the generalized gauge invariance.

Based on Eq.(\ref{eq:master}), we define each morphism of the base abelian category $C^{\nabla}$ to be the non-linear\footnote{Of course, this non-linearity is about the element of $U(\hat{{\mathfrak{g}}})$, and each morphism acts on the object linearly.} transformation operator compatible to the non-linear potential $\aleph$ (namely, where the covariant derivative $\nabla$ is an infinite dimensional vector for this transformation just like the situation such that, in the general theory of relativity, the covariant derivative is a vector on the curved space-time, that is, in our case the parallel section $\Psi^\nabla$ of $\nabla$ such that $\nabla \Psi^\nabla=0$). The objects of $C^\nabla$ are redefined to be compatible to the morphisms and do not need a wave function of the Universe, which is an infinitesimal approximation of the parallel section $\Psi^\nabla$ of $\nabla$.
 {{We change the formulation so that the cohomological contents of wave functions result from the morphisms. In this new vision, the role of the given linear potential $\Psi$ in the $Q$-complexes is substantially taken by the non-linear potential $\aleph$ (in the general theory of relativity, they correspond to Newton potential and the space-time metric respectively), and the category $C^Q$ has only cohomologically trivial contents, that is, the vacuum itself or a unitary factor only.}}
The non-linear potential $\aleph$ describes the dynamics of the morphisms of the derived category $D(C)$ which is the morphism structure of base abelian category $C$. This description is global. Consequently, the non-linear potential $\aleph$ can describe the transition between the stable configurations. In contrast, as explained in the Introduction, the linear wave function $\Psi$ is an infinitesimal local description of $D(C)$ and $\aleph$, and cannot describe the non-perturbative effect nor dynamics of $D(C)$.

This derived category $D(C^\nabla)$ is the conclusive formulation of our model of M-theory vacua via gauged S-duality.
 
\section{Discussion: Cosmic Time and its Complements}
To review the conceptual innovation in our investigation, we discuss the nature of the cosmic time in our model.

In our model, the increment of the cosmic time $\delta\tau$ is defined for any system by the operator such that its expectation value in the system is
\begin{equation}
\la \delta \tau(s)\ra=\biggl[\frac{ k\la Q\ra}{\la \sqrt{(\Omega_0-\Omega)(p^0)}\ra}\biggr](s)\;,\label{eq:timefinal}\end{equation}
for the Casimir operator $\Omega
$ of 
 the affine Lie algebra $\hat{{\mathfrak{g}}}$ in the representation $\Pi(p)$ and its maximum value $\Omega_0$ in the discrete series. The numerator of Eq.(\ref{eq:timefinal}), $k Q(s)$, is proportional to the Hamiltonian ${Q}(s)$. The free part of Eq. (\ref{eq:timefinal}) is proportional to the expectation value of the number of the elements of the system and contains the statistical properties of $\delta\tau(s)$ around the coordinates $s$ as discussed in Section 4.2. (However, we may need a further refinement of Eq.(\ref{eq:timefinal}).)

The way of the variation of the cosmic time is relatively assigned on the Hilbert space $V_s$ of any matter system by Eq.(\ref{eq:Sch}). As already noted, the cosmic time has meaning only as its change in Eq.(\ref{eq:Sch}).
 Our modeling of the cosmic time is completed by incorporating the derived category formulation of the quantum mechanical world in which objects are classified by the retention time of their superpositions. In this formulation, simultaneity is treated as the quasi-equivalence of $Q$-complexes that represent cosmic time evolutions.
 
 The independence of the wave function of the Universe 
\begin{equation}
\Psi[\hat{g},s]\;,
\end{equation}from the cosmic time change was discussed in Section 4.1. The wave function of matter
\begin{equation}
\psi[\hat{g},s;\tau]\;,
\end{equation}
in the Hilbert space $V_s$ depends on the cosmic time changing. The non-triviality of the dependence is measured by the non-triviality of the effect of the statistical variance of the time increment in $V_s$.
 Due to Eq.(\ref{eq:timefinal}), the increment of the cosmic time is affected by the way to represent the wave functions $\psi$.
Via the coordinates $s$, the variation of the cosmic time sums over all of the contributions to its system from the Hilbert spaces, which include spaces to which the wave function of the system $\psi$ does not belong. So, the cosmic time evolves as it is given from outside of the system. Here, we remark that, as explained in Section 3, the spatial expanses are identified with the string probes of them and are also determined by the material probing $3$-vertex states. Thus, the spatial expanses have non-trivial time dependences.

The origin of the scale of the variation of the cosmic time in various systems is an interesting issue. We relate the scale of the variation of the cosmic time with the renormalization by the method of Whitham deformation of the coordinates (and the variables of the representation space), that is, the scales in the Hilbert space.

When we determine the wave function of the Universe $\Psi$ as a function of the coupling constant and all of the coordinates, the functional forms of the increment of the cosmic time and its way to vary would be determined via Eq.(\ref{eq:timefinal}) and Eqs.(\ref{eq:Sch}). In general, it is not actually possible to determine the value of the wave function of the Universe due to the infinite number of data points required. So, we need to rely on a stochastic description, and the change of the cosmic time is a stochastic process for imperfect systems as explained in Section 4.2. Thus, the concept of the increment of the cosmic time for a system $S$ in $V_s$ is also an inevitably imperfect one: that is, the exact form of the effect of the variance of the time increment requires an infinite amount of complementary information from the wave function $\psi$ of $S$ about the variables in the rest of the state space $V_s$ in $V$. Conversely, when the wave function of the Universe $\Psi$ is assumed, the infinite amount of complementary information about the material wave function $\psi$ in the wave function of the Universe $\Psi$ determines the effect of the variance of the time increment of $\psi$ as discussed in Section 4.2. 
 
The quantum mechanically relative interpretation of the concept of time in which every position has its own non-trivial effect of the variance of the time increment is possible due to the fact that the data (coordinates) for determining the increment of the cosmic time are infinitely many and due to the expression of the increment of the cosmic time Eq.(\ref{eq:timefinal}), which sums equally over all of the coordinates.

\section*{Acknowledgements}
I dedicate this article to my mentor Yoshitaka Yamamoto.
\begin{appendix}
\section{Renormalization of the Vacua}
In this appendix, we present a brief account for the renormalization method of the vacua relating to the symmetry and its breaking structures of the vacua.

For a given order parameter $\Lambda$ of a symmetry structure of a vacuum $\Psi$, we define its scale constants as the ratios of the scales of the critical points of the order parameter $\Lambda$ to the bare scale of the vacuum:
\begin{equation}1>\epsilon^{(1)}>\epsilon^{(2)}>\cdots\;,\label{eq:spec}\end{equation}
where the scale of $\epsilon^{(i)}$ gives the critical value $\Lambda^{(i)}$ of the order parameter $\Lambda$.

The renormalization transformation $R$ between the vacuum $\Psi$ and an effective vacuum is formulated by the following multi-phase average method (the Whitham method). It introduces a positive valued relative cut-off scale $\epsilon^{(i)}$ into the effective vacuum $R\Psi$.

 In the Whitham method,\cite{Wh1,Wh2} we redefine our variables of the coordinates $s$ and the variables of the representation space $x$ as the fast variables, which are distinguished from the slow variables $S$ and $X$ that are introduced by the infinite-dimensional vector-valued multi-phase functions $I_s(S)$ and $I_x(X)$ of the wave function, as
 \begin{equation}
1_\omega\cdot s(S)=(\epsilon^{(i)})^{-1} I_s(S)\;,\ \ 1_k\cdot x(X)=(\epsilon^{(i)})^{-1} I_x(X)\;.\label{eq:fast}
\end{equation}

 Then, we deform the additional parameters in the wave function (e.g. the time frequencies $\omega$ and the wave numbers $k$ in the wave function)\begin{equation}a\to A(S,X)\;,\end{equation} by the slow variables $S$ and $X$ as a power series in $\epsilon^{(i)}$:\cite{WKB}\begin{equation}
R\Psi=\sum_{l\in{\boldsymbol{Z}}_{\ge 0}}(\epsilon^{(i)})^{l}\Psi^{(l)}[\hat{g},{s(S)},x(X)|A(S,X)]\;,\label{eq:WKB}
\end{equation} satisfying the equation \begin{equation}({{Q}}R)\Psi=0\;,\label{eq:Whitham}\end{equation}
with BRST charge ${Q}$. Here, we impose the symmetries, whose order parameters $\Lambda$ are still zero at the critical scale $\epsilon^{(i)}$, on $R\Psi$. The periods of the fast variables $s(S)$ and $x(X)$ of $\Psi^{(0)}$ are constant in the slow variables $S$ and $X$. As an additional condition, to exclude the secular terms in $\Psi^{(l_+)}$ for $l_+=1,2,\ldots$, we set the periods of any wave function $\Psi^{(l_+)}$ in the fast variables $s$ and $x$ to be the same as the periods of $\Psi^{(0)}$ in them.
 With these deformed solutions $\Psi^{(l)}$ for $l=0,1,2,\ldots$, we average the fast variables $s$ and $x$ and regard $R\Psi$ as functions of the modulus parameter $\hat{g}$ and the slow variables $S$ and $X$. By Eq.(\ref{eq:Whitham}), the dependence of the effective vacuum $R\Psi$ on the modulus parameter $\hat{g}$ is specified.

The consistency conditions on the $\Psi^{(0)}$ and $\Psi^{(l_+)}$ parts of the Whitham deformation, found by the comparisons of both sides of the Kugo-Ojima physical state condition at each power of $(\epsilon^{(i)})^{l}$ for $l=0,1,2,\ldots$, are the integrability conditions on the multi-phase functions $I_s(S)$ and $I_x(X)$ and an infinite number of local conservation laws under the averages of the contributions from the fast variables $s$ and $x$ respectively.\cite{Wh1,Wh2,WKB} They are just Eq.(\ref{eq:Whitham}). This is why we adopt the Whitham method as the renormalization.
\end{appendix}

\end{document}